 \documentclass[preprint,prd,amssymb,preprintnumbers,superscriptaddress,nofootinbib]{revtex4-1}
 \usepackage{array}
 \usepackage{graphicx}
\usepackage{amsfonts}
\usepackage{amssymb}
\usepackage{amsthm}
\usepackage{amsmath}
\usepackage{multirow}
\usepackage{color}
\usepackage{float}
\usepackage{geometry}
\usepackage{hyperref}
\usepackage{cleveref}

\newcommand{\beq}{\begin{equation}}
\newcommand{\eeq}{\end{equation}}
\newcommand{\bea}{\begin{eqnarray}}
\newcommand{\eea}{\end{eqnarray}}

\newcommand{\gsim}{\lower.7ex\hbox{$\;\stackrel{\textstyle>}{\sim}\;$}}
\newcommand{\lsim}{\lower.7ex\hbox{$\;\stackrel{\textstyle<}{\sim}\;$}}
\newcommand{\be}{\begin{equation}}
\newcommand{\ee}{\end{equation}}
\newcommand{\ba}{\begin{eqnarray}}
\newcommand{\ea}{\end{eqnarray}}

\usepackage{caption}

\begin{document}

\title{Modular Family Symmetry in Fluxed GUTs}

\author{Vasileios Basiouris}
\email{v.basiouris@uoi.gr}
\affiliation{Physics Department, University of Ioannina, 45110, Ioannina, Greece}

\author{Miguel Crispim Rom\~ao}
\email{miguel.romao@durham.ac.uk}
\affiliation{Institute for Particle Physics Phenomenology, Department of Physics, Durham University, Durham DH1 3LE, U.K.}
\affiliation{School of Physics and Astronomy, University of Southampton,Southampton SO17 1BJ, United Kingdom}
\affiliation{Laborat\'orio de Instrumenta\c{c}\~ao e F\'isica Experimental de Part\'iculas, Escola de Ciências, Departamento de F\'isica, Universidade do Minho, Portugal}

\author{Stephen F. King}
\email{s.f.king@soton.ac.uk}
\affiliation{School of Physics and Astronomy, University of Southampton,Southampton SO17 1BJ, United Kingdom}

\author{George K. Leontaris}
\email{leonta@uoi.gr}
\affiliation{Physics Department, University of Ioannina, 45110, Ioannina, Greece}

\begin{abstract}
	\noindent
	We discuss modular family symmetry in effective theories based on generic properties of bottom-up local F-theory inspired GUTs broken by fluxes, which we refer to as Fluxed GUTs. We argue that the Yukawa couplings will depend on the complex structure moduli of the matter curves in such a way that they can be modular forms associated with these symmetries.
	To illustrate the approach we analyse in detail a concrete local fluxed $SU(5)$ GUT with modular $S_4$ family symmetry.
\end{abstract}

\preprint{IPPP/24/42}

\maketitle

\section{Introduction}

The predictive power of discrete flavour symmetries in fermion mass hierarchies has long been emphasised. In attempting to explain the fermion mass hierarchies and mixings, various non-abelian discrete groups were successfully applied as family symmetries to numerous extensions of the standard model (see e.g. \cite{King:2013eh} and references therein).

Such family symmetry must be spontaneously broken by vacuum expectation values (VEVs) of scalar flavon fields, leading to the vacuum alignment problem, which leads to additional complications. Modular invariance was suggested to overcome such complications~\cite{Feruglio:2017spp}. In the simplest examples, the only flavon is a complex modulus $\tau$ that transforms under the modular group $SL(2, \mathbb{Z})$. Finite modular symmetries $\Gamma_N$ can emerge as the quotient group of $SL(2, \mathbb{Z})$ with one of its principal congruence subgroups $\Gamma(N)$ of level $N$. Yukawa couplings are represented by modular forms which transform under the finite level $N$ quotient group, $\Gamma_N$, that further restricts the possible mass textures and increases the predictive power of the effective theory. Subsequently, such finite modular family symmetry has been widely studied from the top-down and bottom-up perspectives
(see e.g.~\cite{Kobayashi:2023zzc,Ding:2023htn} for recent reviews).

From the top-down perspective, there have been attempts to investigate such symmetries in the context of heterotic superstring motivated models~\cite{Nilles:2021glx}. Indeed, upon orbifold compactification of the six extra dimensions, the modular group and its congruence group emerge naturally in the four-dimensional effective theory. The origin of the modular group, in particular, is attributed to the specific geometry of the compactification manifold.

Another example is an effective supergravity model from Type IIB compactification~\cite{Kobayashi:2020hoc,Kikuchi:2023clx}. Its massless spectrum contains several moduli which correspond to deformations of the Calabi-Yau (CY) manifold. Among them, the complex structure, the K\"ahler and the axio-dilaton, the latter being a complex scalar combination $\tau$ where the real part is the RR axion $C_0$ and its imaginary is the inverse of the string coupling $g_s^{-1}=e^{-\phi}$ where $\phi$ is the dilaton field. Taking into account that the massless spectrum is described by an associated supergravity and the fact that $g_s$ takes only positive values, the field $\tau$ admits values subject to $SL(2,\mathbb{Z})$, which is nothing else than S-duality.

Several quantities and parameters of the effective theory, such as the K\"ahler potential, the superpotential, and Yukawa couplings, depend on the field $\tau$ and have well-defined transformation properties under (some congruence group of) $SL(2,\mathbb{Z})$. Analogously, in toroidal compactifications, there are discrete symmetries, dubbed T-dualities, which relate Type IIB theory with compactification radius $R$ to Type IIA associated with the inverse radius $1/R$.  Constraints similar to the above mentioned are also expected to be imposed on the effective theory~(see~\cite{Ferrara:1989qb} and review~\cite{DHoker:2022dxx}).

In this paper,  motivated by the rich framework of local F-Theory~\cite{Beasley:2008dc,Beasley:2008kw},
we extend these ideas to bottom-up effective theories based on generic properties of local F-theory inspired
GUTs broken by fluxes that share important features and similarities to explicit F-Theory models, which we refer to as Fluxed GUTs~\cite{CrispimRomao:2017tnd}.
Yukawa couplings in F-Theory emerge as integrals of overlapping wavefunctions around the intersection of matter curves, each a Riemannian surface. In this construction, the wavefunctions are solutions to the equations of motion of the compactified Yang-Mills (YM) theory on the 7-branes, with an undetermined holomorphic dependency on the coordinates of the matter curve. This observation leads us to conclude that the structure of the Yukawa couplings will depend on the complex structure moduli of the matter curves in such a way that if these curves exhibit modular symmetries, then the Yukawa couplings can be modular forms associated with these symmetries.

Following from this observation, we develop a bottom-up approach based on Fluxed GUTs with discrete modular symmetry, where the fields on chosen matter curves, subject to specified fluxes, transform as some representation of a discrete modular family group and carry assigned modular weights. Yukawa couplings, arising from the overlapping wavefunction integrals at matter curve intersections, are then promoted to modular forms of appropriate weight to balance the Yukawa operator.
As a first example of such a bottom-up approach, we have analysed in detail a concrete local fluxed $SU(5)$ GUT with modular $S_4$ family symmetry. We show that, although the model fits the data very well, with the best point having vanishing $\chi^2$, it is over-parameterised and therefore is not predictive. Nevertheless, it provides some insight into fermion mass hierarchies and serves as a demonstration of the bottom-up approach to modular family symmetry in F-Theory inspired Fluxed GUTs.

This work is organised as follows. In~\cref{sec:modular_symmetry} we revise  modular symmetry invariant field theories to set notation and definitions. In~\cref{sec:origin_modular_f_theory} we review how Yukawa couplings are obtained in local F-Theory.
In~\cref{sec:modular-f-gut} we argue that
such Yukawa couplings should be modular forms since they inherit modular symmetry properties from the geometric moduli, which parametrise the matter field wavefunctions.
We then present a concrete F-Theory inspired Fluxed $SU(5)$ GUT with $S_4$ modular symmetry, including a fit to experimental data in~\cref{sec:numerics}. In~\cref{sec:conclusions} we conclude and provide an outlook for our framework.

\section{Modular Symmetry Invariant Supersymmetric Field Theories\label{sec:modular_symmetry}}

In this section, we introduce modular forms of even weight, following~\cite{Feruglio:2017spp}  to set the notation and definitions that we will be using for the rest of this work. We first define the homogeneous modular group, $\Gamma$, is composed of $2\times2$ matrices such that
\begin{equation}
	\Gamma = SL(2,\mathbb{Z}) = \left\{\begin{pmatrix}a & b\\c & d \end{pmatrix} \Big| \ a,b,c,d \in \mathbb{Z}, \ a d - b c = 1 \right\} \ .
\end{equation}

$\Gamma$ has two generators, $S$ and $T$, which respect
\begin{align}
	S^4 = (S T)^3 = I,
\end{align}
where $I$ is the identity element, and can be represented in matrix form as
\begin{align}
	S = \begin{pmatrix}
		    0 & 1 \\ -1 & 0
	    \end{pmatrix}
	,\
	T = \begin{pmatrix}
		    1 & 1 \\ 0 & 1
	    \end{pmatrix} \ .
\end{align}

Another closely related group is the inhomogeneous modular group, $\bar \Gamma$, which is the group of linear fraction transformations, $\gamma$, acting on the upper half complex plane, $\mathcal{H} = \{\tau \in \mathbb{C}, \ \Im (\tau)>0\}$, as
\begin{align}
	\tau \mapsto \gamma \tau = \frac{a \tau +b}{c \tau +d}, \ \text{s.t. } a,b,c,d \in \mathbb{Z}, \ a d - bc =1 \ .
\end{align}
Since the elements $\pm \gamma$ produce the same linear fractional transformation, we can see that this group is related to $\Gamma$ as
\begin{align}
	\bar \Gamma = \Gamma / \{\pm I\} = \left\{\begin{pmatrix}a & b\\c & d \end{pmatrix}/\{\pm \mathbb{I}\} \Big| \ a,b,c,d \in \mathbb{Z}, \ a d - b c = 1\right\} ,
\end{align}
where $\mathbb{I}$ is the identity $2\times2$ matrix, and we note that it is isomorphic to the projective matrix group $PSL(2,\mathbb{Z})\cong SL(2,\mathbb{Z}) / \{\pm I\}$. $\bar \Gamma$ is also generated by $S$ and $T$, but since $I$ and $-I$ are identified, we have $S^2 = I$, and they act on a complex number $\tau \in \mathcal{H}$ as
\begin{align}
	S: \tau \mapsto - \frac{1}{\tau} , \ T : \tau \mapsto \tau + 1
\end{align}

An important sequence of subgroups of $\Gamma$ are the principal congruence subgroups of level $N$, $\Gamma(N)$, which are given by
\begin{align}
	\Gamma(N) = \left\{\begin{pmatrix}a & b\\c & d \end{pmatrix} \in \Gamma \Big | \ \begin{pmatrix}a & b\\c & d \end{pmatrix} = \begin{pmatrix}1 & 0\\0 & 1 \end{pmatrix} \mod N \right\} \ ,
\end{align}
which are infinite normal subgroups of $\Gamma$. Since the congruence of unity are all the integers, we have $\Gamma(1)=\Gamma$. Furthermore, we define $\bar \Gamma(N) = \Gamma(N)/\{\pm I\}$ for $N=1,2$, and since $-I$ is not an element of $\Gamma(N)$ for $N>2$ we have $\bar \Gamma (N) = \Gamma(N)$ for $N>2$.

The quotients $\Gamma_N = \bar \Gamma / \bar \Gamma(N)$ are called inhomogeneous finite modular groups of level $N$. They are finite and discrete, and are generated by $S$ and $T$  respecting
\begin{align}
	S^2 = (ST)^3 = T^N = I \ .
\end{align}

Inhomogeneous finite modular groups $\Gamma_N$ have been studied in great detail in the literature~\cite{Feruglio:2017spp,Kobayashi:2018vbk,Yao:2020qyy,Qu:2021jdy,Ding:2019xna}, with $\Gamma_2 \cong S_3$, $\Gamma_3 \cong A_4$, $\Gamma_4 \cong S_4$, and $\Gamma_5 \cong A_5$, as well as higher levels~\cite{Li:2021buv,Ding:2020msi}. For a full, comprehensive, list see~\cite{Ding:2023htn} and references therein.

An important aspect of modular symmetries is the notion of modular forms that are holomorphic functions of a complex parameter $\tau\in\mathcal{H}$, which we will take to be the modulus of a compact space, with special transformation properties. A modular form $f(\tau)$, of weight $k$ and level $N$ is a holomorphic function of $\tau \in \mathcal{H}$ that under the action of $\gamma \in \bar\Gamma(N)$ transforms as
\begin{align}
	f(\gamma \tau) = (c \tau + d)^k f(\tau), \ \forall \gamma \in \bar\Gamma (N) \ ,
\end{align}
and it can been shown~\cite{diamond2005first} that, for even $k$, these span a finite dimension linear space, $\mathcal{M}_k(\bar\Gamma(N))$, and in which it is possible to find a basis such that the modular forms furnish a representation of the inhomogeneous finite modular group $\Gamma_N$
\begin{align}
	f_r (\gamma\tau) = (c \tau +d)^k \rho_r(\gamma)f_r(\tau), \ \forall \gamma \in \bar \Gamma \ ,
\end{align}
where $\gamma$ is a representative element of $\Gamma_N$, $\rho_r(\gamma)$ is the representation matrix of $\gamma$ in the representation $r$ of $\Gamma_N$.

We will now use these to build supersymmetric theories with modular invariance, in order to construct a framework for model building. We will follow~\cite{Feruglio:2017spp,King:2020qaj} and provide the uniform notation and definitions that we will use for the remainder of this work.

An $\mathcal{N}=1$ supersymmetric theory is defined by two functions: the K{\"a}hler potential, $K(\Phi_I,\bar\Phi_I,\tau,\bar\tau)$ - a real function of the chiral superfields $\{\Phi_I\}$ and the moduli $\{\tau\}$ - and the superpotential $W(\Phi,\tau)$ - a holomorphic function of the same superfields. We will consider the case where there is only one modulus field.

The superpotential admits the generic expansion
\begin{align}
	W(\Phi_I,\tau) = \sum_n Y_{I_1,\dots,I_n}\Phi_{I_1}\dots\Phi_{I_n} \ ,
\end{align}
where the couplings $Y_{I_1,\dots,I_n}$ are modular forms of weight $k_Y$, and will then transform under the action of $\bar\Gamma$. The chiral superfields, $\{\Phi_I\}$, will also transform non-trivially, but are otherwise not modular forms. Under the action of modular group, we have
\begin{align}
	 & \tau        \mapsto \gamma \tau = \frac{a \tau + b}{c \tau + d}              \\
	 & \Phi_I       \mapsto (c \tau +d)^{-k_I}\rho_I(\gamma)\Phi_I                  \\
	 & Y_{I_1,\dots,I_n} \mapsto (c \tau + d)^{k_Y}\rho_Y(\gamma) Y_{I_1,\dots,I_n}
\end{align}
where $\gamma$ is an element of $\bar\Gamma$, and $\rho_I(\gamma)$ and $\rho_Y(\gamma)$ are representation matrices of $\gamma$ in the finite modular subgroup, $\Gamma_{N}$, $-k_I$ and $k_Y$ are modular weights. Therefore, superpotential level invariance is obtained when the sum of the modular weights for each coupling vanish, and if there is a singlet in the product of representations
\begin{align}
	k_Y        & = k_{I_1} + \dots k_{I_n}                      \\
	\mathbf{1} & \subset \rho_Y \rho_{I_1} \dots \rho_{I_n} \ .
\end{align}

When considering the full supergravity theory, the relevant defining quantity which needs to be invariant is the K\"ahler function, $G(\Phi_I,\bar\Phi_I,\tau,\bar\tau)$, given by
\begin{align}
	G(\Phi_I,\bar\Phi_I,\tau,\bar\tau) = K(\Phi_I,\bar\Phi_I,\tau,\bar\tau) + \ln W(\Phi_I,\tau) + \ln \bar W(\bar\Phi_I,\bar\tau) \ .
\end{align}
This allows for the sum of modular weights to be non-vanishing at the superpotential level, as long as the leftover is absorbed by the K{\"a}hler potential through a K{\"a}hler transformation. However, to do so one usually needs an explicit form of the K{\"a}hler potential. For example, if one considers a K{\"a}hler potential with the customary no-scale structure with a chiral superfield expansion
\begin{align}
	K(\Phi_I,\bar\Phi_I,\tau,\bar\tau) = - h \ln(-i (\tau - \bar\tau)) + \sum_I (-i(\tau-\bar\tau))^{-k_I} |\Phi_I|^2 \ ,
\end{align}
with $h$ a positive number, the superpotential now transforms as
\begin{align}
	W(\Phi,\tau) \mapsto (c\tau+d)^{-h}W(\Phi,\tau) \ .
\end{align}
This effectively amounts to change the condition for vanishing modular weights to
\begin{align}
	k_Y     = k_{I_1} + \dots k_{I_n} - h \ ,
\end{align}
whilst each superpotential coupling still needs to be an $\Gamma_{N}$ invariant singlet.

\section{Yukawa Couplings in F-Theory\label{sec:origin_modular_f_theory}}

In this section, we briefly review Yukawa couplings in local F-theory constructions.
In order to define the Yukawa couplings in fluxed compactifications, our starting point is an effective F-Theory GUT model, which is derived from an ADE-type singularity with the world-volume of a 7-brane that wraps the space $R^{3,1}\times {S}$ with ${S}$ being a K\"ahler manifold of two complex dimensions $z_1,z_2$. At low energies, F-Theory is described by an eight-dimensional YM theory on $R^{3,1}\times \cal{S}$ which must be topologically twisted to preserve $N=1$ supersymmetry.

The compactification space is a fibred eight-dimensional space (CY fourfold $CY_4$) where the fibre over the base $B_3=CY_3$ associated with the six-dimensional compact space is described by a two-dimensional torus whose modulus is the axio-dilaton $\tau=C_0+ie^{-\phi}= C_0+i/g_s$.  Therefore, the $SL(2,\mathbb{Z})_\tau$ S-duality describes the variation of the modulus $\tau$ of the 2-torus over the compactification manifold. The geometric configuration consists of 7-branes filling the Minkowski 4D-space while wrapping a 4D `surface'  $S$ -- associated with some GUT symmetry -- which is a complex K\"ahler manifold so that supersymmetry is preserved.
The four-dimensional effective F-Theory model arises upon compactification of the eight-dimensional theory on  $S$.
The possible GUT groups, in particular, are associated with specific types of geometric singularities
where the modulus $\tau$ acquires certain values. The massless fields of the low energy spectrum reside on Riemann surfaces, called matter curves, formed by 7-branes intersecting the GUT surface, while Yukawa couplings are formed at specific points where triple intersections of matter curves occur.~\footnote{Equivalently, the torus over $B_3$ can be described by the Weierstra\'ss equation $y^2=x^3 +f(z) x^2 +g(z)$, where $z$ is a coordinate of the complex projective space $CP^1$ (Riemann sphere). Then $\tau =\tau(z) =C_0(z)+ie^{-\phi(z)}$ and singularities occur at $\Delta(z_i)=0$. The torus is associated with the invariant $j(\tau(z))$, which, together with the vanishing of $\Delta$ determines $\tau \sim \frac{1}{2\pi i}\log(z-z_i)$. Hence, for given $z$, $\tau$ is fixed. Also, going around the singularity, there is a shift to the real part of the modulus $C_0\to C_0+1$ that corresponds to $\tau \to \tau+1$ of $SL(2, \mathbb{Z})_\tau$.}

Within this framework,  the corresponding gauge theory is that of the eight-dimensional $N=1$ supersymmetric YM theory with minimal field content. The bosonic spectrum, in particular, includes the gauge field $A$ and a holomorphic two-form scalar $\Phi$.  Both fields are found in the adjoint representation and descend from the decomposition of the 10-dimensional gauge field ${\cal A}$.  Since 7-branes are wrapped on a curved $R^{(3,1)}\times S$ space, unbroken ${\cal N}=1$ supersymmetry requires $\Phi$ to be a holomorphic (2,0)-form as a result of the topological twisting~\cite{Vafa:1994tf}.
\footnote{Within such an F-Theory framework it is
	well known that there are many complex structure moduli, associated with the positions of the 7-branes. The positions of the 7-branes are determined by tuning the complex structure moduli and can produce additional structure in the elliptic fibration
	\cite{Dierigl:2020lai}.}

The superpotential $W_{8d}$ of the eight-dimensional fields
and an associated D-term take the form
\be
W_{8d}= m_*^4 \int_S{\rm Tr}(F\wedge \Phi),\; \; D=\int_S\omega\wedge F+\frac 12 [\Phi,\bar\Phi ]~,\label{W8d}
\ee
where $F=dA-iA\wedge A$, and $\omega =i g/2 (dz_1\wedge  d \bar z_1+dz_2\wedge  d \bar z_2)$ is the K\"ahler form on $S$.

The eight-dimensional fields can be organised as one ${\cal N}=1$ vector multiplet, ${\bf V}$, and two ${\cal N}=1$ chiral supermultiplets, ${\bf A}_{\bar m}$ and ${\bf\Phi}_{mn}$,
\begin{align}
	{\bf V}          & =
	(A_{\mu},\eta^{\alpha},{\cal D})                                           \\
	{\bf A}_{\bar m} & =(A_{\bar m},\psi_{\bar m}^{\alpha},{\cal G}_{\bar m})) \\
	{\bf\Phi}_{mn}   & =(\varphi_{mn},\chi_{mn}^{\alpha},{\cal H}_{ mn}) \ ,
	\label{multi1}
\end{align}
where
${\cal G}_{\bar m}, {\cal H}_{\bar m\bar n}$ are  $F$-term components, and ${\cal D}$ represents the $D$-term, whilst,
$\eta^{\dot\alpha}, \psi_{\bar m}^{\alpha},\chi_{mn}^{\dot\alpha}$
are the fermionic components, which, in the twisted YM theory  are associated with a zero, one- and two-form respectively~\cite{Beasley:2008dc},
\[
	\eta^{\dot\alpha},\;\psi^{\alpha}=\psi_{\bar m}^{\alpha}d\bar z^{\bar m},\;
	\chi^{\dot\alpha}=\bar\chi_{\bar m\bar n}^{\dot\alpha}d\bar z^{\bar m}\wedge d\bar z^n~.
\]
The indices $m,n$ take the values 1,2, the complex scalars $A_{\bar m},\varphi_{mn}$ have dimensions of mass $M$ and ${\cal G},{\cal H},{\cal D}$ of squared mass $M^2$.

To preserve the supersymmetric vacuum, all variations of the eight-dimensional fields must vanish. In the context  of the four-dimensional theory, this corresponds to imposing the F- and D-flantess of the superpotential. Minimising the
superpotential (\ref{W8d}) and imposing D-flatness, one arrives
at the following equations
\begin{align}
	\bar \partial_A \Phi                              & = 0   \\
	F^{(2,0)}                                         & = 0   \\
	\omega \wedge F + \frac{1}{2}[\Phi^\dagger, \Phi] & = 0~.
\end{align}
The above equations have long been derived in reference~\cite{Vafa:1994tf} and are the basic ingredients for studying the properties of fields in generic  7-brane configurations.
Here, we are interested in solutions for massless fields residing on 7-brane configurations. The equations can be solved
by expanding the fields	$ A,\Phi$ assuming linear fluctuations around the background:
\begin{align}
	A_{\bar m}  \to\langle A_{\bar m} \rangle+ a_{\bar m},\;\; &
	\;\;\Phi       \to\langle \Phi\rangle + \varphi~,
\end{align}
with the definitions
\begin{align}
	a=a_{\bar z_1}d{\bar z_1}+a_{\bar z_2}d{\bar z_2},\;\; & \;\;\varphi= \varphi_{\bar z_1\bar z_2}d{\bar z_1}\wedge d{\bar z_2} \ .
\end{align}

\noindent
Then, keeping only linear terms regarding the fluctuations $\varphi, a$, in the holomorphic gauge the EoM take the form
\begin{align}
	\bar \partial_{\langle A\rangle} a                                                     & = 0   \label{e1}  \\
	\bar \partial_{\langle A\rangle} \varphi - i [a, \langle \bar\Phi\rangle]              & =0                \\
	\omega\wedge \partial_{\langle A\rangle} a -\frac 12 [\langle \bar\Phi\rangle,\varphi] & = 0\label{e3} \ .
\end{align}

Substituting the expansions of the fields into~(\ref{W8d})  it is found that
the holomorphic trilinear Yukawa coupling is  written in terms of $\phi$ and $ a$ as follows
\begin{equation}
	W_{\rm Yuk} = - i m_*^4\int_S \text{Tr}(\varphi \wedge a \wedge a)\label{YuInt} \ ,
\end{equation}
where $m_*$ is the scale associated with the supergravity limit of F-Theory.

The fluctuations $\varphi$ and $a$ can be
determined by solving the equations (\ref{e1}-\ref{e3}) for a variety of diagonal or non-diagonal backgrounds, the latter being known as T-branes~\cite{Cecotti:2009zf}. They are associated with the zero-modes residing on the matter curves and when three of the latter  define  triple intersection a Yukawa coupling is formed. Depending on the details of the model, it is often the case that multiple zero-modes are accommodated on the same matter curve.

It can be shown that the general form of the solution for zero modes localised on a specific matter curve, say $z_2$ , takes the generic form

\be
\varphi = R_a \chi_a =R_af(z_2) g(z_{1},\bar z_{1},q) e^{-\sqrt{ M_{z_1}^4+m^4} z_{1}\bar z_{1}} e^{\pm 2 M_{z_2}^2z_2 \bar z_2} \ ,
\label{WFsol}
\ee
where $M_{z_i}$ appear when fluxes are also introduced~\footnote{For
	example, in a $U(3)$ model the flux assumes the form $\langle F\rangle =-(2i/3) M^2 (\bar z_1\wedge dz_1-\bar z_2\wedge dz_2){\rm diag}(1,-2,1)$. In a generic context, however, when non-Abelian T-branes are considered, a non-primitive flux is required $\omega\wedge F\ne 0$ to satisfy the D-term.
	For a comprehensive presentation, see review \cite{Font:2015slq}.}

It can be observed that locally the solution is described by a Gaussian profile, with its peak along the matter curve and waning out along the transverse direction $z_1$.
The function $f(z_2)$ is a holomorphic function of $z_2$ left undetermined from the equations of motion
and $R_a$ encodes the group structure
associated with the background. Analogous solutions can be written for the other intersecting matter curves in the vicinity of the triple intersection. The integration (\ref{YuInt})~~\footnote{In section 3 eq~(3.28) of~\cite{Cecotti:2009zf}- using the notion of twisted one-forms, the connection $\psi= a+\hat{\kappa}\wedge \varphi,\, \hat{\kappa}=g^{1\bar i}g^{2\bar j}\bar{\Omega}_{\overline{ijz}}d\bar z$ is implemented -
and the Yukawa coupling receives a symmetric form. See the appendix for the relevant computation.} is performed over the three overlapping wavefunctions where all of them are peaked at the triple intersection and since they are strongly localised, the integral can be restricted to a small region near the intersection point. At every triple intersection  the gauge symmetry is enhanced and generically  zero-mode states are assembled into  representations of the  higher symmetry. At the same time, multiple states accommodated on a  certain matter curve may be organised into representations of the underlying symmetry of
the complex structure  of the matter curve.

We would like to note that, while the above discussion makes use of the local limit for computing the Yukawa couplings, we expect that if the integrals were to be computed using a complete global model, the dependency of the stabilised complex structure moduli, which parametrise the positions of the 7-branes and therefore the shape and the geometry of the matter curves~\cite{Dierigl:2020lai}, would remain in the computed Yukawa coupling. Another approach where this happens is in simplified global toy models where the Yukawa couplings are shown to be dependent on the holomorphic sections of the spectral cover equation, which are themselves functions of the complex structure moduli~\cite{Cvetic:2019sgs}. Additionally, we note that the in Type-IIB string theory one can explicitly obtain background vacua that transform non-trivially under congruence subgroups of $SL(2,\mathbb{Z})$, breaking the symmetry down to a discrete subgroup~\cite{Kobayashi:2020hoc}. Given that Type-IIB string theory is the perturbative limit of F-Theory, it is reasonable to assume that similar vacua exist in F-Theory in such a way that matter supported at matter curves and the Yukawa couplings arising from their intersections will transform non-trivially under the surviving discrete modular symmetry.

We discuss now the overall dependence of the Yukawa couplings on the mass scales of the theory, and their relation to the axio-dilaton  modulus~\cite{Beasley:2008kw}. In string frame, the overall scale $m_*$ in (\ref{YuInt}), is given by $m_*^8=m_s^8 g_s^{-2}$~\cite{Beasley:2008kw}, hence, the resulting dependence of the Yukawa coupling on $g_s$ is (see details in section 4 of~\cite{Beasley:2008kw})
\begin{equation}
	\lambda \propto \frac{m_*^4}{m_s^4} = \frac{1}{g_s} \ .
\end{equation}
The string coupling is related to the GUT scale. Indeed, let ${\cal V}_S\sim R_S^4\sim 1/m^4_{GUT}$ be the volume of the GUT surface and ${\cal V}_B\sim R^6$ that of the base $B_3$ of the fibration. Compactification to four dimensions implies $M_{Pl}^2\approx m_*^8 {\cal V}_B$ while from the kinetic term of the field strength it follows $\alpha_{GUT}^{-1}\approx m_*^4{\cal V}_S$.
Combining these relations we obtain a rough estimate
\be
\frac{1}{\alpha_{GUT}}\sim {m_*}^4{\cal V}_S= \frac{m_*^4}{m_s^4}\frac{m_s^4}{m_{GUT}^4}=\frac{m_s^4}{m_{\small GUT}^4}\frac{1}{g_s} \ ,
\ee

Taking into account the various normalisation effects, the dependence on $g_s$ is more involved. One finds
\begin{equation}
	\lambda =C\, a_{GUT}^{3/4} \ ,
\end{equation}
where $C$ may depend on other moduli fields via the wavefunctions of the form~(\ref{WFsol}) involved in the triple intersections.
This implies that in the effective theory  S-duality symmetry is broken by the Yukawa lagrangian, unless the undetermined parts of the wavefunctions associated with the Yukawa coupling under consideration exhibit the appropriate dependence on $g_s$.

\section{Fluxed GUTs and Finite Discrete Modular Symmetry\label{sec:modular-f-gut}}

In the previous section we reviewed the basic F-Theory approach to Yukawa couplings and presented a generic solution for the EoM.
Now we shall discuss a possible connection between bottom-up  Fluxed GUTs in local F-theory models and the emergence of discrete modular flavour symmetries in such geometries. The key point lies in the role played by the complex structure moduli $z_i$, which could in principle parametrize the K{\"a}hler manifold of the F-theory's geometric configuration. Generalized fluxes could modify the background leading to symmetry breaking of the internal modular symmetry to a residual flavor symmetry, where this effect is manifested in the wavefunction's holomorphic function. Consequently, Yukawa couplings
should be modular forms via their computation through evaluating the integral of intersecting wavefunctions.

From the above argument we infer that the Yukawa couplings inherit group properties encoded in the matter wavefunctions. The latter depend on the complex structure moduli through holomorphic functions of the complex coordinates $z_i$ which are left unspecified by the EoM. In the present context of a generic phenomenological approach the explicit knowledge of the fluxes formulation in local F-theory fluxed backgrounds does not need to be specified. The key point is that fluxes magnetising the 7-branes are expected to preserve some symmetry of the complex structure sector, since the fluxed superpotential of the moduli fields is subject to modular restrictions for preserving a supersymmetric vacuum. Therefore, if the Yukawa sector for the ordinary matter of the superpotential is required to retain the same modular symmetry or a subgroup thereof, its origin is expected to come from the unspecified part of that solution. Based on the above reasoning, we assume that the wavefunctions transform as modular forms as follows,
\begin{equation}
	f(\tau_i) \to (c \tau_i + d)^{-k_i} f(\tau_i) \ ,
\end{equation}
where $\tau_i$ is a complex structure modulus associated with the complex coordinate $z_i$. The holomorphic Yukawa coupling, arising from the intersection of three matter curves, will then transform as a modular form in a non-trivial representation of the congruence subgroup of the modular group, as discussed in
section~\ref{sec:modular_symmetry}.

To illustrate the ideas just outlined of local F-Theory inspired Fluxed GUTs with modular symmetry, we now present a concrete example of a bottom-up model sharing important results with F-Theory models. Let us consider an $SU(5)$ GUT embedded in $E_6$ which has been derived in an F-Theory framework~\cite{Callaghan:2011jj}. The novel feature of this paradigm will be the inclusion of an $S_4$ finite modular family symmetry, which is isomorphic to the inhomogeneous modular
group $\Gamma_4\cong S_4$.\footnote{We note that $S_4$ is a choice for the example model discussed herein. Additionally, we will assume even modular weights as it is common in bottom-up model building. Many explicit string constructions lead to half-integer modular weights, which would need the extension to the metaplectic group. We leave these constructions for future work.}

As has already been pointed out, in this context matter fields reside on curves which are  formed at the intersections of 7-branes with the GUT surface $S$, itself wrapped with 7-branes. We consider a divisor with an $E_{6}$ geometric singularity which, according to the F-Theory prescription, corresponds to an $E_6$ gauge symmetry of the effective theory.
In the present setup, there are three matter curves accommodating three $27_{t'_i}$ representations of $E_6$. These are distinguished from each other by the weights $t'_i$ of the $SU(3)$ Cartan sublagebra $(t'_1+t'_2+t'_3=0)$.
We impose a ${\cal Z}_2$ monodromy $t_1'\leftrightarrow t_2'$, and hence only two distinct matter curves remain, e.g. $\Sigma_{27_{t_{1,3}'}}$, and use $U(1)$ fluxes to reduce the gauge symmetry down to $SU(5)$. Alternatively, one may derive this model starting from the maximum admissible (well behaved) singularity that corresponds to a $E_8$ gauge symmetry subsequently decomposed to
\begin{equation}
	E_8\supset SU(5)\times SU(5)_{\perp}\supset SU(5)\times U(1)^4_{t_i},\;\; \sum_{i=1}^5t_i=0~,
\end{equation}
where now $t_i$ correspond to the Cartan subalgebra of $SU(5)_{\perp}$. The $E_6$ and $SU(5)\times SU(5)_{\perp}$ properties of the matter and Higgs  multiplets are given in Table \ref{tab1}. Due to the aforementioned restrictions on $t_i'$ and the monodromy imposed, the only  allowed trilinear $E_6$ term  in the superpotential is $W\supset 27_{t_{1}'}27_{t_{1}'}27_{t_{3}'}$.  We then assign the fermion supermultiplets to $27_{t'_{1}}$ and the Higgs fields to $27_{t'_{3}}$.

	{\renewcommand{\arraystretch}{1.1}
		\begin{table}[]
			\makebox[\textwidth][c]{
				\begin{tabular}{cccc}
					\hline \hline
					$E_6$       & $SO(10)$ & $SU(5)$          & Weight     \\
					\hline
					$27_{t_1'}$ & $16$     & $\overline{5}_3$ & $t_1+t_5$  \\

					$27_{t_1'}$ & $16$     & $10_M$           & $t_1$      \\

					$27_{t_1'}$ & $16$     & $\theta_{15}$    & $t_1-t_5$  \\

					$27_{t_1'}$ & $10$     & $5_1$            & $-t_1-t_3$ \\

					$27_{t_1'}$ & $10$     & $\overline{5}_2$ & $t_1+t_4$  \\

					$27_{t_1'}$ & $1$      & $\theta_{14}$    & $t_1-t_4$  \\

					$27_{t_3'}$ & $16$     & $\overline{5}_5$ & $t_3+t_5$  \\

					$27_{t_3'}$ & $16$     & $10_2$           & $t_3$      \\

					$27_{t_3'}$ & $16$     & $\theta_{35}$    & $t_3-t_5$  \\

					$27_{t_3'}$ & $10$     & $5_{H_u}$        & $-2t_1$    \\

					$27_{t_3'}$ & $10$     & $\overline{5}_4$ & $t_3+t_4$  \\

					$27_{t_3'}$ & $1$      & $\theta_{34}$    & $t_3-t_4$  \\
					\hline\hline
				\end{tabular}
			}
			\caption{\small $SO(10)$ and $SU(5)$ decompositions of $27\in E_6$. The $SU(5)$ indices in $5_i, 10_j$ representations  designate their origin of the corresponding matter curve ($\Sigma_{5_i}$ and $\Sigma_{10_j}$),  and $10_M$ accommodates ordinary matter fields.}
			\label{tab1}
		\end{table}}%
We break the $SU(5)$ gauge symmetry by turning on a flux along $U(1)_Y\in SU(5)$, which also splits the $10$ and $\bar{5}$ representations of $SU(5)$.
However, anomaly cancellation conditions impose constraints on the multiplicities of the latter which are as follows:
\begin{align}
	M_{10_M}=M_{5_1}=-M_{5_2}=-M_{5_3}, \; M_{10_2}=-M_{5_4}=-M_{5_5}=M_{5_{H_u}}.
\end{align}
Furthermore, to eliminate extraneous and exotic matter derived from the decomposition of the 78-dimensional representation, we impose the conditions
\begin{align}
	M_{10_3}=M_{10_4}=M_{5_6}=N_8=N_9=0,
\end{align}
which imply that~\cite{Callaghan:2011jj}
\begin{align}
	\tilde{N}\equiv N_7~.
\end{align}

The SM zero mode states derived from the complete $27_{t'_i}$ representations after various successive symmetry-breaking stages with the $U(1)$ fluxes are shown in the last column of Table~\ref{no1}. Their multiplicities are expressed in terms of the flux integers which have remained undetermined by the consistency conditions mentioned above.
	{\renewcommand{\arraystretch}{1.1}
		\begin{table}[]
			\makebox[\textwidth][c]{
				\begin{tabular}{ccccccc}
					\hline\hline
					$E_6$       & $SO(10)$ & $SU(5)$          & Weight vector & $N_Y$        & $M_{U(1)}$    & SM particle content \\
					\hline
					$27_{t_1'}$ & $16$     & $\overline{5}_3$ & $t_1+t_5$     & $\tilde{N}$  & $-M_{5_3}$    &
					$-M_{5_3}d^c+(-M_{5_3}+\tilde{N})L$                                                                            \\

					$27_{t_1'}$ & $16$     & $10_M$           & $t_1$         & $-\tilde{N}$ & $-M_{5_3}$    &
					$-M_{5_3}Q+(-M_{5_3}+\tilde{N})u^c+(-M_{5_3}-\tilde{N})e^c$                                                    \\

					$27_{t_1'}$ & $16$     & $\theta_{15}$    & $t_1-t_5$     & $0$          & $-M_{5_3}$    &
					$-M_{5_3}\nu^c$                                                                                                \\

					$27_{t_1'}$ & $10$     & $5_1$            & $-t_1-t_3$    & $-\tilde{N}$ & $-M_{5_3}$    &
					$-M_{5_3}D+(-M_{5_3}-\tilde{N})H_u$                                                                            \\

					$27_{t_1'}$ & $10$     & $\overline{5}_2$ & $t_1+t_4$     & $\tilde{N}$  & $-M_{5_3}$    &
					$-M_{5_3}\overline{D}+(-M_{5_3}+\tilde{N})H_d$                                                                 \\

					$27_{t_1'}$ & $1$      & $\theta_{14}$    & $t_1-t_4$     & $0$          & $-M_{5_3}$    &
					$-M_{5_3}S$                                                                                                    \\

					$27_{t_3'}$ & $16$     & $\overline{5}_5$ & $t_3+t_5$     & $-\tilde{N}$ & $M_{5_{H_u}}$ &
					$M_{5_{H_u}}d^c+(M_{5_{H_u}}-\tilde{N})L$                                                                      \\

					$27_{t_3'}$ & $16$     & $10_2$           & $t_3$         & $\tilde{N}$  & $M_{5_{H_u}}$ &
					$M_{5_{H_u}}Q+(M_{5_{H_u}}-\tilde{N})u^c+(M_{5_{H_u}}+\tilde{N})e^c$                                           \\

					$27_{t_3'}$ & $16$     & $\theta_{35}$    & $t_3-t_5$     & $0$          & $M_{5_{H_u}}$ &
					$M_{5_{H_u}}\nu^c$                                                                                             \\

					$27_{t_3'}$ & $10$     & $5_{H_u}$        & $-2t_1$       & $\tilde{N}$  & $M_{5_{H_u}}$ &
					$M_{5_{H_u}}D+(M_{5_{H_u}}+\tilde{N})H_u$                                                                      \\

					$27_{t_3'}$ & $10$     & $\overline{5}_4$ & $t_3+t_4$     & $-\tilde{N}$ & $M_{5_{H_u}}$ &
					$M_{5_{H_u}}\overline{D}+(M_{5_{H_u}}-\tilde{N})H_d$                                                           \\

					$27_{t_3'}$ & $1$      & $\theta_{34}$    & $t_3-t_4$     & $0$          & $M_{5_{H_u}}$ &
					$M_{5_{H_u}}S$                                                                                                 \\
					\hline\hline
				\end{tabular}
			}
			\caption{\small Complete $27$s of $E_6$ and their $SO(10)$ and $SU(5)$ decompositions.
				The indices of the $SU(5)$ non-trivial states $10,5$ refer to the labelling of the corresponding matter curve (we use the notation of~\cite{Dudas:2010zb}). We impose the extra conditions on the integers $N_Y$ and $M_{U(1)}$ from the requirement of having complete 27s of $E_6$ and no 78 matter. The $SU(5)$ matter states decompose into SM states as $\overline{5}\rightarrow d^c,L$ and $10\rightarrow Q,u^c,e^c$ with right-handed neutrinos $1\rightarrow \nu^c$, while the $SU(5)$ Higgs states decompose as $5\rightarrow D,H_u$ and $\overline{5}\rightarrow \overline{D},H_d$, where $D, \overline{D}$ are exotic colour triplets and antitriplets. We identify RH neutrinos as $\nu^c=\theta_{15,35}$ and extra singlets from the 27 as $S=\theta_{14,34}$.}\label{no1}
		\end{table}}%

In the present work, an explicit model is constructed by choosing the fluxes given in~\cref{tab:fluxes}.
{\renewcommand{\arraystretch}{1.1}
\begin{table}[]
	\makebox[\textwidth][c]{
		\begin{tabular}{cccccccccc}
			\hline \hline
			$M_{10_M}$ & $M_{5_3}$ & $M_{5_1}$ & $M_{5_2}$ & $ M_{10_2}$ & $M_{5_{5}}$ & $M_{5_4}$ & $M_{H_{u}}$ & $M_{\theta_{15}}$ & $\tilde{N}$ \\
			\hline
			4          & $-4$      & 3         & $-3$      & $-1$        & 1           & 0         & 0           & 2                 & 1           \\
			\hline\hline
		\end{tabular}
	}
	\caption{\small  The choice of Fluxes used in this model.\label{tab:fluxes}}
\end{table}}
This choice leads to the spectrum given in~\cref{tab:f-theory-spectrum} where both the down quarks and leptons originate from $27_{t_{1}'}$. We note that this choice of fluxes provides a solution to the doublet-triplet splitting problem by allowing for light Higgs doublets, while the coloured counterparts acquire masses of the scale of the compactification, thereby suppressing coloured triplet-mediated proton decay.
	{\renewcommand{\arraystretch}{1.1}
		\begin{table}[]
			\makebox[\textwidth][c]{
				\begin{tabular}{cccccccc}
					\hline\hline
					$E_6$                          & $SO(10)$           & $SU(5)$          & Weight vector & $N_Y$ & $M_{U(1)}$ & SM particle content & Low energy spectrum \\
					\hline
					$27_{t_1'}$                    & $16$               & $\overline{5}_3$ & $t_1+t_5$     & $1$   & $4$        &
					$4d^c+5L$                      & $3d^{c}+3L$                                                                                                            \\

					$27_{t_1'}$                    & $16$               & $10_M$           & $t_1$         & $-1$  & $4$        &
					$4Q+5u^c+3e^c$                 & $3Q+3u^{c}+3e^{c}$                                                                                                     \\

					$27_{t_1'}$                    & $16$               & $\theta_{15}$    & $t_1-t_5$     & $0$   & $3$        &
					$3\nu^c$                       & -                                                                                                                      \\

					$27_{t_1'}$                    & $10$               & $5_1$            & $-t_1-t_3$    & $-1$  & $3$        &
					$3D+2H_u$                      & -                                                                                                                      \\

					$27_{t_1'}$                    & $10$               & $\overline{5}_2$ & $t_1+t_4$     & $1$   & $3$        &
					$3\overline{D}+4H_d$           & $H_{d}$                                                                                                                \\

					$27_{t_3'}$                    & $16$               & $\overline{5}_5$ & $t_3+t_5$     & $-1$  & $-1$       &
					$\overline{d^c}+2\overline{L}$ & -                                                                                                                      \\

					$27_{t_3'}$                    & $16$               & $10_2$           & $t_3$         & $1$   & $-1$       &
					$\overline{Q}+2\bar{u^c}$      & -                                                                                                                      \\

					$27_{t_3'}$                    & $16$               & $\theta_{35}$    & $t_3-t_5$     & $0$   & $0$        &
					$-$                            & -                                                                                                                      \\

					$27_{t_3'}$                    & $10$               & $5_{H_u}$        & $-2t_1$       & $1$   & $0$        &
					$H_u$                          & $H_{u}$                                                                                                                \\

					$27_{t_3'}$                    & $10$               & $\overline{5}_4$ & $t_3+t_4$     & $-1$  & $0$        &
					$\overline{H_d}$               & -                                                                                                                      \\

					$27_{t_3'}$                    & $1$                & $\theta_{34}$    & $t_3-t_4$     & $0$   & $1$        &
					$\theta_{34}$                  & -                                                                                                                      \\

					-                              & $1$                & $\theta_{31}$    & $t_3-t_1$     & $0$   & $4$        &
					$\theta_{31}$                  & -                                                                                                                      \\

					-                              & $1$                & $\theta_{53}$    & $t_5-t_3$     & $0$   & $1$        &
					$\theta_{53}$                  & -                                                                                                                      \\

					-                              & $1$                & $\theta_{14}$    & $t_1-t_4$     & $0$   & $3$        &
					$\theta_{14}$                  & -                                                                                                                      \\

					-                              & $1$                & $\theta_{45}$    & $t_4-t_5$     & $0$   & $2$        &
					$\theta_{45}$                  & -                                                                                                                      \\
					\hline\hline
				\end{tabular}
			}
			\caption{\small Complete $27$s of $E_6$ and their $SO(10)$ and $SU(5)$ decompositions. We use the notation of ref~\cite{Dudas:2010zb} for the indices of the $SU(5)$ states and impose the extra conditions on the integers $N_Y$ and $M_{U(1)}$ from the requirement of having complete 27s of $E_6$ and no 78 matter. The $SU(5)$ matter states decompose into SM states as $\overline{5}\rightarrow d^c,L$ and $10\rightarrow Q,u^c,e^c$ with right-handed neutrinos $1\rightarrow \nu^c$, while the $SU(5)$ Higgs states decompose as $5\rightarrow D,H_u$ and $\overline{5}\rightarrow \overline{D},H_d$, where $D, \overline{D}$ are exotic colour triplets and antitriplets. We identify RH neutrinos as $\nu^c=\theta_{15}$. Extra singlets are needed to given mass to neutrinos and exotics and to ensure F- and D- flatness.}
			\label{tab:f-theory-spectrum}
		\end{table}}%

As we have argued in the previous section, the states supported on a matter curve will inherit modular symmetry properties related to the complex structure moduli parametrising that curve. Therefore, states supported on a given curve are expected to have the same modular weights and to furnish full representations of the discrete modular group that survives the compactification. Imposing these modular symmetry properties in the above representations, a version of the model presented above with non-trivial discrete modular group $S_4$ can be written as
\begin{align}
	\mathcal{W} & =  \alpha \; (u^c_{1,2} Q_{1,2}  Y_1^{(4)})_1 H_u + \beta \; (u^c_{1,2} Q_{1,2} Y_2^{(4)})_1 H_u +\gamma \; (u^c_{3} Q_{3} Y_{1}^{(4)})_1  H_u +\delta \; (u^c_{1,2} Q_{3}  Y_{2}^{(4)})_1  H_u\notag                         \\
	            & +\bigg(\alpha^{\prime}\; (d^c_{1,2}  Q_{1,2} Y_1^{(6)})_1  H_d + \beta^{\prime}\; (d^c_{1,2} Q_{1,2} Y_{2}^{(6)})_1 H_d +\gamma_1^{\prime}\; (d^c_{3} Q_{1,2}  Y_{2,1}^{(8)})_1 H_d  \notag                                   \\
	            & \qquad + \gamma^{\prime}\; (d^c_{3} Q_{1,2} Y_{2,2}^{(8)})_1  H_d+  \delta^{\prime}\; (d^c_{1,2} Q_{3} Y_2^{(6)})_1 H_d +\epsilon^{\prime}\; (d^c_{3} Q_{3} Y_1^{(8)})_1  H_d  \bigg)\dfrac{\theta_{31}}{M},\label{superpot1}
\end{align}
where $M$ is the F-Theory characteristic compacfication scale and, for simplicity, we will set  ${\theta_{31}}/{M}\simeq 1$ as we expect the VEVs of the singlets to be close to the scale $M$ and this quantity can be reabsorbed into the definition of the primed coefficients.

	{\renewcommand{\arraystretch}{1.1}
		\begin{table}[h]
			\makebox[\textwidth][c]{
				\begin{tabular}{ccccc}
					\hline\hline
					MSSM fields                    & Matter Curves & Charge    & $S_4$ & k   \\
					\hline
					$Q_{1,2},u^c_{1,2},e^c_{1,2} $ & $10_M$        & $t_1$     & $2$   & $2$ \\
					$Q_{3},u^c_{3},e^c_{3} $       & $10_M$        & $t_1$     & $1$   & $2$ \\
					$ d^c_{1,2}, L_{1,2}  $        & $\bar{5}_3$   & $t_1+t_5$ & $2$   & $4$ \\
					$d^c_{3}, L_{3}$               & $\bar{5}_3$   & $t_1+t_5$ & $1$   & $6$ \\
					$H_u$                          & $5_{H_u}$     & $-2t_1$   & $1$   & $0$ \\
					$H_d$                          & $5_2$         & $t_1+t_4$ & $1$   & $0$ \\
					$\nu^c$                        & $\theta_{15}$ & $t_1-t_5$ & $3$   & $0$ \\
					\hline\hline
				\end{tabular}
			}
			\caption{\small Perpendicular charges, modular weights, and $S_4$ discrete modular group representations associated with the matter curves hosting the model from~\cref{tab:f-theory-spectrum}.\label{tab:model_modular_props}}
		\end{table}}

\noindent
According to the superpotential \eqref{superpot1}, the up-type quarks Yukawa matrix is given by
\begin{align}
	\lambda_u=\begin{pmatrix}
		          \alpha  \left(Y_1^2+Y_2^2\right)-\beta  \left(Y_2^2-Y_1^2\right) & 2 \beta  Y_1 Y_2                                                 & \delta  \left(Y_2^2-Y_1^2\right) \\
		          2 \beta  Y_1 Y_2                                                 & \alpha  \left(Y_1^2+Y_2^2\right)+\beta  \left(Y_2^2-Y_1^2\right) & 2 \delta  Y_1 Y_2                \\
		          0                                                                & 0                                                                & \gamma  \left(Y_1^2+Y_2^2\right)
	          \end{pmatrix}~,
	\label{eq:Lambdau}
\end{align}
and for the down-type quarks, the relevant Yukawa matrix is written as
\begin{align}
	\lambda_d=\left(\begin{smallmatrix}
			                \alpha^{\prime}  Y_1 \left(3 Y_2^2-Y_1^2\right)-\beta^{\prime}  Y_1 \left(Y_1^2+Y_2^2\right) & \beta^{\prime}  Y_2 \left(Y_1^2+Y_2^2\right) & \delta^{\prime}  Y_1 \left(Y_1^2+Y_2^2\right) \\
			                \beta^{\prime}  Y_2 \left(Y_1^2+Y_2^2\right) & \alpha^{\prime}  Y_1 \left(3 Y_2^2-Y_1^2\right)+\beta^{\prime}  Y_1 \left(Y_1^2+Y_2^2\right) & \delta^{\prime}  Y_2 \left(Y_1^2+Y_2^2\right) \\
			                \gamma^{\prime}  \left(Y_1^2-3 Y_2^2\right) Y_1^2+\gamma _1^{\prime} \left(Y_2^2-Y_1^2\right) \left(Y_1^2+Y_2^2\right) & \gamma^{\prime}  Y_1 Y_2 \left(Y_1^2-3 Y_2^2\right)+2 \gamma _1^{\prime} Y_1 Y_2 \left(Y_1^2+Y_2^2\right) & \epsilon^{\prime} \left(Y_1^2+Y_2^2\right){}^2
		                \end{smallmatrix}\right) \ .
	\label{eq:Lambdad}
\end{align}

\noindent
The charged leptons have the same Yukawa matrix structure as the down-type quarks. However, inspecting the spectrum of F-Theory zero modes in~\cref{tab:f-theory-spectrum}, we see that the three families of $L$, $Q$, $e^c$, and $d^c$ descend from different linear combinations of UV states from F-Theory zero modes. Therefore, the superpotential coefficients for the down-type quarks and the charged leptons are not the same, leading to a realisation of a Georgi-Jarlskog mechanism~\cite{Georgi:1979df}. We then write down the charged leptons Yukawa matrix as
\begin{align}
	\lambda_L=\left(\begin{smallmatrix}
			                \alpha^{\prime\prime}  Y_1 \left(3 Y_2^2-Y_1^2\right)-\beta^{\prime\prime}  Y_1 \left(Y_1^2+Y_2^2\right) & \beta^{\prime\prime}  Y_2 \left(Y_1^2+Y_2^2\right) & \delta^{\prime\prime}  Y_1 \left(Y_1^2+Y_2^2\right) \\
			                \beta^{\prime\prime}  Y_2 \left(Y_1^2+Y_2^2\right) & \alpha^{\prime\prime}  Y_1 \left(3 Y_2^2-Y_1^2\right)+\beta^{\prime\prime}  Y_1 \left(Y_1^2+Y_2^2\right) & \delta^{\prime\prime}  Y_2 \left(Y_1^2+Y_2^2\right) \\
			                \gamma^{\prime\prime}  \left(Y_1^2-3 Y_2^2\right) Y_1^2+\gamma _1^{\prime\prime} \left(Y_2^2-Y_1^2\right) \left(Y_1^2+Y_2^2\right) & \gamma^{\prime\prime}  Y_1 Y_2 \left(Y_1^2-3 Y_2^2\right)+2 \gamma _1^{\prime\prime} Y_1 Y_2 \left(Y_1^2+Y_2^2\right) & \epsilon^{\prime\prime} \left(Y_1^2+Y_2^2\right){}^2
		                \end{smallmatrix}\right) \ ,
	\label{eq:Lambdal}
\end{align}
where the modular form components have the same dependence on $\tau_d$  as those appearing in the down-type quark Yukawa matrix.

In the following discussion, we are going to sketch a scenario in which the conjugate right-handed neutrinos are identified with the singlets $\theta_{15}$, which are included in the particle spectrum of the F-Theory model. Since these fields are considered as degrees of freedom residing in the transverse space of the matter curves \cite{Bouchard:2009bu}, this fact leads us to consider the case that they do not carry any modular weight. However, a simple model is presented here in which the singlets transform as a triplet under the $S_4$ modular symmetry. In addition to the singlets mentioned before, more degrees of freedom are needed to give a Majorana mass to $\theta_{15}$, leading to the implementation of a (type-I) seesaw scenario for the light neutrino masses. An additional constraint that must be taken into account, is that the additional singlets of the model have to cancel the perpendicular charges of the coupling.

\noindent
The superpotential, following the transformation properties of Table \eqref{tab:model_modular_props}, is written as:

\begin{equation}
	\mathcal{W}_{\nu}= \zeta \;  (\nu^c L_{1,2} Y^{(4)}_3)_1  H_u + \eta  \; (\nu^c  L_{3}  Y^{(6)}_3)_1 H_u + \lambda\;  (\nu^c \nu^c)_1 \; \dfrac{\theta_{53}^2\theta_{31}^2}{M^3},
	\label{SupNu}
\end{equation}
The first two couplings generate the following Yukawa matrix  for the neutrino Dirac mass:
\begin{align}
	\lambda_\nu=\begin{pmatrix}
		            -2\zeta Y_2 Y_3                                & 0                                                   & \eta Y_1(Y_4^2-Y_5^2)            \\
		            -\dfrac{1}{2}\zeta (\sqrt{3}Y_1 Y_4 +Y_2 Y_5)  & \dfrac{\sqrt{3}}{2} \zeta (\sqrt{3}Y_1 Y_5+Y_2 Y_4) & -\eta Y_3(Y_1Y_4+\sqrt{3}Y_2Y_5) \\
		            -\dfrac{1}{2}  \zeta (\sqrt{3}Y_1 Y_5+Y_2 Y_4) & \dfrac{\sqrt{3}}{2}\zeta (\sqrt{3}Y_1 Y_4 +Y_2 Y_5) & \eta Y_3(Y_1Y_5+\sqrt{3}Y_2Y_4)
	            \end{pmatrix}~,
	\label{eq:lambdan}
\end{align}
where the modular form components depend on the same modulus of the up-type quark, $\tau_u$.
The last coupling of~(\ref{SupNu}) provides  Majorana masses to  the right-handed neutrinos. The corresponding  Majorana mass matrix is
\begin{align}
	M_{R}=\begin{pmatrix}
		      1 & 0 & 0 \\
		      0 & 0 & 1 \\
		      0 & 1 & 0
	      \end{pmatrix}\lambda \dfrac{\theta_{53}^2\theta_{31}^2}{M^3},
	\label{eq:majorana}
\end{align}
where we will take $\lambda \theta_{53}^2\theta_{31}^2/M^3= \tilde \lambda M_{GUT}$.

Next  we  implement a type-I seesaw mechanism to determine the light neutrino mass matrix  given by
\begin{align}
	M_{\nu}=-M_D^T M_R^{-1} M_D~,
\end{align}
where $M_D = v \lambda_\nu$, with $v=173$ GeV being the Standard Model Higgs vacuum expectation value.

It is worth emphasising that the model described in this section represents a significant improvement to the models hitherto previously derived from F-Theory as the finite modular symmetry of the model presented above allows explicit Yukawa matrices to be constructed and analysed. This is in contrast to other typical F-Theory inspired models, such as those presented in~\cite{Callaghan:2011jj},
which involve rank one Yukawa matrices and must appeal to uncontrolled non-pertubative flux effects in order to obtain phenomeonologically acceptable structures.
Additionally, F-theory motivated Fluxed GUTs typically predict extra vector-like states, unlike
other typical bottom-up modular models purely based on modular symmetry without further dynamical assumptions ~\cite{Ding:2023htn}. Although the vector-like states in the model presented in~\cref{tab:f-theory-spectrum} are assumed to be heavy with large Dirac masses being allowed after compactification, in other constructions they can be light~\cite{CrispimRomao:2017tnd} under appropriate modifications of the string parameters such as fluxes, leading to testable predictions.

\section{Numerical study\label{sec:numerics}}
In this section we perform a brief numerical study to find whether the model presented in the previous section and explicitly stated by the superpotential in~\cref{superpot1} can provide a good fit to quark masses and mixing. To do this, we will compare the model predictions against the values of the quark masses and mixing data at the GUT scale, which, for $\tan \beta=5$, can be found in~\cref{tab:quarkdata}. The neutrino data are taken from the latest \texttt{NuFit 5.3},~\cite{Esteban:2020cvm} and is shown in~\cref{tab:leptondata} alongside the charged lepton Yukawa eigenvalues.

	{\renewcommand{\arraystretch}{1.1}
		\begin{table}[h!]
			\makebox[\textwidth][c]{
				\begin{tabular}{cccc}
					\hline\hline
					\multicolumn{4}{c}{\textbf{Quark and CKM Data}}                                           \\
					\hline
					$y_d$ & $(4.81\pm 1.06) \times 10^{-6}$  & $\theta_{12}$ & $13.027^\circ\pm 0.0814^\circ$ \\
					$y_s$ & $(9.52\pm 1.03) \times 10^{-5}$  & $\theta_{23}$ & $2.054^\circ\pm 0.384^\circ$   \\
					$y_b$ & $(6.95\pm 0.175) \times 10^{-3}$ & $\theta_{13}$ & $0.1802^\circ\pm 0.0281^\circ$ \\
					$y_u$ & $(2.92\pm 1.81) \times 10^{-6}$  & $\delta_{CP}$ & $69.21^\circ\pm 6.19^\circ$    \\
					$y_c$ & $(1.43\pm 0.100) \times 10^{-3}$ &               &                                \\
					$y_t$ & $0.534\pm 0.0341$                &               &                                \\
					\hline\hline
				\end{tabular}
			}
			\caption{\small Quark and CKM data~\cite{Antusch:2013jca,Bjorkeroth:2015ora,Okada:2020rjb}, where the quoted uncertainties are 1-$\sigma$.}
			\label{tab:quarkdata}
		\end{table}
	}

	{\renewcommand{\arraystretch}{1.1}
		\begin{table}[h!]
			\makebox[\textwidth][c]{
				\begin{tabular}{cccc}
					\hline\hline
					\multicolumn{4}{c}{\textbf{Lepton and PMNS Data}}                                                                            \\
					\hline
					$y_e$             & $(1.97\pm 0.024) \times 10^{-6}$               & $\sin^2\theta^L_{12}$ & $0.307\pm 0.012$                \\
					$y_\mu$           & $(4.16\pm 0.05) \times 10^{-4}$                & $\sin^2\theta^L_{23}$ & $0.572\pm 0.023$                \\
					$y_\tau$          & $(7.07\pm 0.073) \times 10^{-3}$               & $\sin^2\theta^L_{13}$ & $(2.203\pm 0.58)\times 10^{-2}$ \\
					$\Delta m^2_{12}$ & $(7.41\pm 0.21) \times 10^{-5} \text{ eV}^2$   & $\delta^L_{CP}$       & $197^\circ\pm 41^\circ$         \\
					$\Delta m^2_{13}$ & $(2.511\pm 0.027) \times 10^{-3} \text{ eV}^2$ &                       &                                 \\
					\hline\hline
				\end{tabular}
			}
			\caption{\small Lepton and PMNS data. Neutrino masses are given in normal ordering~\cite{Antusch:2013jca,Bjorkeroth:2015ora,Okada:2020rjb,Esteban:2020cvm}. The quoted uncertainties are 1-$\sigma$ and asymmetrical uncertainty intervals were made symmetric by choosing the larger value, so that a  Gaussian likelihood profile can be used in the numerical analysis.}
			\label{tab:leptondata}
		\end{table}}

We use the effective Yukawa coupling matrices for the quarks,~\cref{eq:Lambdau,eq:Lambdad}, as well as for the neutrinos,~\cref{eq:lambdan,eq:majorana}, to compute the predictions and compare them to the data in~\cref{tab:quarkdata,tab:leptondata}. Although the coefficients of the superpotential are in principle calculable in F-Theory (see, for example,~\cite{Heckman:2008qa,Leontaris:2010zd} for Yukawa couplings and \cite{CrispimRomao:2015naq} for R-Parity violating terms), in this work we will consider these coefficients as free parameters and leave the study of their computation for future work. Additionally, we also have the dependency on the complex structure moduli fields parametrising the geometry of the matter curves, from which the matter fields inherit their discrete modular symmetry properties. Since up- and down-type quark Yukawas emerge at different intersection points in the internal geometry between different curves, the geometry describing each Yukawa coupling is in general different from each other and parametrised by its own modulus, i.e. the components of the modular forms appearing in the up- and down-type Yukawas can depend on different moduli fields, $\tau_u$ and $\tau_d$, respectively. However, the charged leptons (neutrino) Yukawa matrix arises from the same intersection as the down-type (up-type) quark Yukawas and should therefore depend on the same modulus. Therefore, our (effective) parametric freedom encompasses:
\begin{itemize}
	\item Four complex coefficients ($\alpha$, $\beta$, $\delta$, $\gamma$) and a complex modulus ($\tau_u$) for the up-type Yukawa matrix,
	\item three complex coefficients ($\zeta$, $\eta$, $\tilde \lambda$) for the neutrino sector (as well as a dependency on $\tau_u$),
	\item six coefficients ($\alpha^\prime$, $\beta^\prime$, $\gamma^\prime$, $\gamma^\prime_1$, $\epsilon^\prime$) and one complex modulus ($\tau_d$) for the down-type Yukawa matrix,
	\item six coefficients ($\alpha^{\prime\prime}$, $\beta^{\prime\prime}$, $\gamma^{\prime\prime}$, $\gamma^{\prime\prime}_1$, $\epsilon^{\prime\prime}$) for the charged lepton Yukawa matrix (as well as a dependency on $\tau_d$).
\end{itemize}
This sums up to a total of 19 complex parameters, or 38 real parameters. Although this seems to over-parameterise our problem, as we only have 19 observables in~\cref{tab:quarkdata,tab:leptondata}, we must reiterate that the complex coefficients are in principle calculable in F-Theory and that the analysis present here simplifies this step.

To find whether we can jointly fit all observables, we employ an artificial intelligence search algorithm called Covariant Matrix Approximation Evolutionary Strategy (CMAES)~\cite{cmaes}, which was first proposed in~\cite{deSouza:2022uhk} to simplify the task of finding valid points in highly constrained multidimensional BSM parameter spaces.\footnote{See also~\cite{Romao:2024gjx} for a recent application to the $Z_3$ 3HDM, where CMAES was shown to have up to nine orders of magnitude improvement in sampling efficiency over random sampling.} CMAES can be seen as a population-based optimisation algorithm that can find minima of any arbitrary function, irrespective of its continuity and differentiability. Therefore, we will use CMAES to minimise the minus log-likelihood of the data, $D$, given a point of the parameter space, $\theta$,\footnote{Or, equivalently, to minimise the sum of the $\chi^2$.}
\begin{equation}
	- llh(D | \theta) = \sum_i \frac{(\bar\mu_i - \mu_i(\theta))^2}{2 \sigma_i^2} \ ,
	\label{eq:llh}
\end{equation}
where $i$ runs over the observables, $\mu_i(\theta)$ is the prediction for the observable $i$ given a parameter space point $\theta$, the data, $D$, are comprised of the set of tuples $\{(\bar\mu_i, \sigma_i)\}$, where $\bar\mu_i$, $\sigma_i$ are, respectively, the central and 1-$\sigma$ uncertainty values of the observables and are listed in~\cref{tab:quarkdata,tab:leptondata}, and we have assumed a Gaussian  profile likelihood for the data. We implemented CMAES using the \texttt{python} package \texttt{cmaes}~\cite{nomura2024cmaes}, and we performed 1000 independent runs, each running until converged to a minimum of~\cref{eq:llh}, and kept all points whose observable predictions were within 3-$\sigma$.\footnote{This methodology is justified by the fact that our goal is not to draw a complete portrait of the parameter space, but rather to find examples of viable points.}. The parameters of our model were bounded, so that the superpotential coefficients remain perturbative and the moduli take values in their fundamental domain with an upper bound on the imaginary part
\begin{equation}
	\left\{ \tau_i \in \mathbb{C},\ s.t.\ |\Re{(\tau_i)}| \leq 0.5 \wedge \sqrt{1 - \Re{(\tau_i)}^2  }  \leq \Im{(\tau_i)} \leq 10  \right\} \ , i=u,\ d\ \ .
	\label{eq:domain}
\end{equation}

Multiple successful runs converged, generating $18\times10^6$ points that fit all observables within 3-$\sigma$. The best point across all runs, that minimises the~\cref{eq:llh} at a value $1.15\times10^{-15}$ (i.e., effectively with vanishing $\chi^2$ or likelihood of $1$), is given by the set of parameters (to up to one decimal digit)
\begin{align}
	\left(\alpha, \beta, \delta, \gamma \right)=                                                                                      & (-1.8 \times 10^{-3} + 1.8 \times 10^{-5}i, 4.5 \times 10^{-5}  -1.4 \times 10^{-4}i, \nonumber  \\
	                                                                                                                                  & 3.2 \times 10^{-4} + 1.8 \times 10^{-3}i, 1.8 \times 10^{-1} + 4.0 \times 10^{-2}i) \nonumber    \\
	\left(\lambda^\prime,\beta^\prime,\gamma^\prime,\gamma_1^\prime,\epsilon^\prime\right)  =                                         & (2.1 \times 10^{-5}  -8.8 \times 10^{-8}i, -3.3 \times 10^{-5} + 3.2 \times 10^{-8}i, \nonumber  \\
	                                                                                                                                  & -4.4 \times 10^{-5} + 7.2 \times 10^{-5}i, -2.3 \times 10^{-4}  -2.3 \times 10^{-4}i, \nonumber  \\
	                                                                                                                                  & -7.7 \times 10^{-5} + 1.4 \times 10^{-4}i, 1.3 \times 10^{-4}  -4.6 \times 10^{-3}i) \nonumber   \\
	(\zeta, \eta, \tilde \lambda) =                                                                                                   & (-6.1 \times 10^{-2} + 9.1 \times 10^{-1}i, -1.5 \times 10^{-1} + 7.2 \times 10^{-3}i, \nonumber \\
	                                                                                                                                  & 1.8 \times 10^{-1} + 5.3 \times 10^{-2}i) \nonumber                                              \\
	\left(\lambda^{\prime\prime},\beta^{\prime\prime},\gamma^{\prime\prime},\gamma_1^{\prime\prime},\epsilon^{\prime\prime}\right)  = & (7.5 \times 10^{-5} + 1.4 \times 10^{-7}i, 2.7 \times 10^{-4}  -7.3 \times 10^{-7}i,\nonumber    \\
	                                                                                                                                  & -1.0 \times 10^{-3} + 1.9 \times 10^{-3}i, 1.2 \times 10^{-4} + 2.6 \times 10^{-8}i,\nonumber    \\
	                                                                                                                                  & -2.9 \times 10^{-3} + 4.1 \times 10^{-5}i, 4.2 \times 10^{-4} -1.7 \times 10^{-3}i)\nonumber     \\
	\tau_u  =                                                                                                                         & -4.1 \times 10^{-1} + 9.1 \times 10^{-1}i\nonumber                                               \\
	\tau_d  =                                                                                                                         & -5.0 \times 10^{-1} + 1.2 i\ ,
	\nonumber                                                                                                                                                                                                                            \\
	\label{eq:best_point}
\end{align}
where we organised the parameters by mass sector. We notice that the point above requires some hierarchy between superpotential coefficients which should be around the same order, e.g. $|\gamma|\sim\mathcal{O}(1)$ whereas $|\alpha|\sim\mathcal{O}(10^{-3})$. This hierarchy between coefficients of operators arising from the intersection of the same matter curves at the same intersection point is at odds with our F-Theory expectations, which requires further study involving their explicit computation.

\begin{figure}[H]
	\centering
	\includegraphics[width=0.7\textwidth]{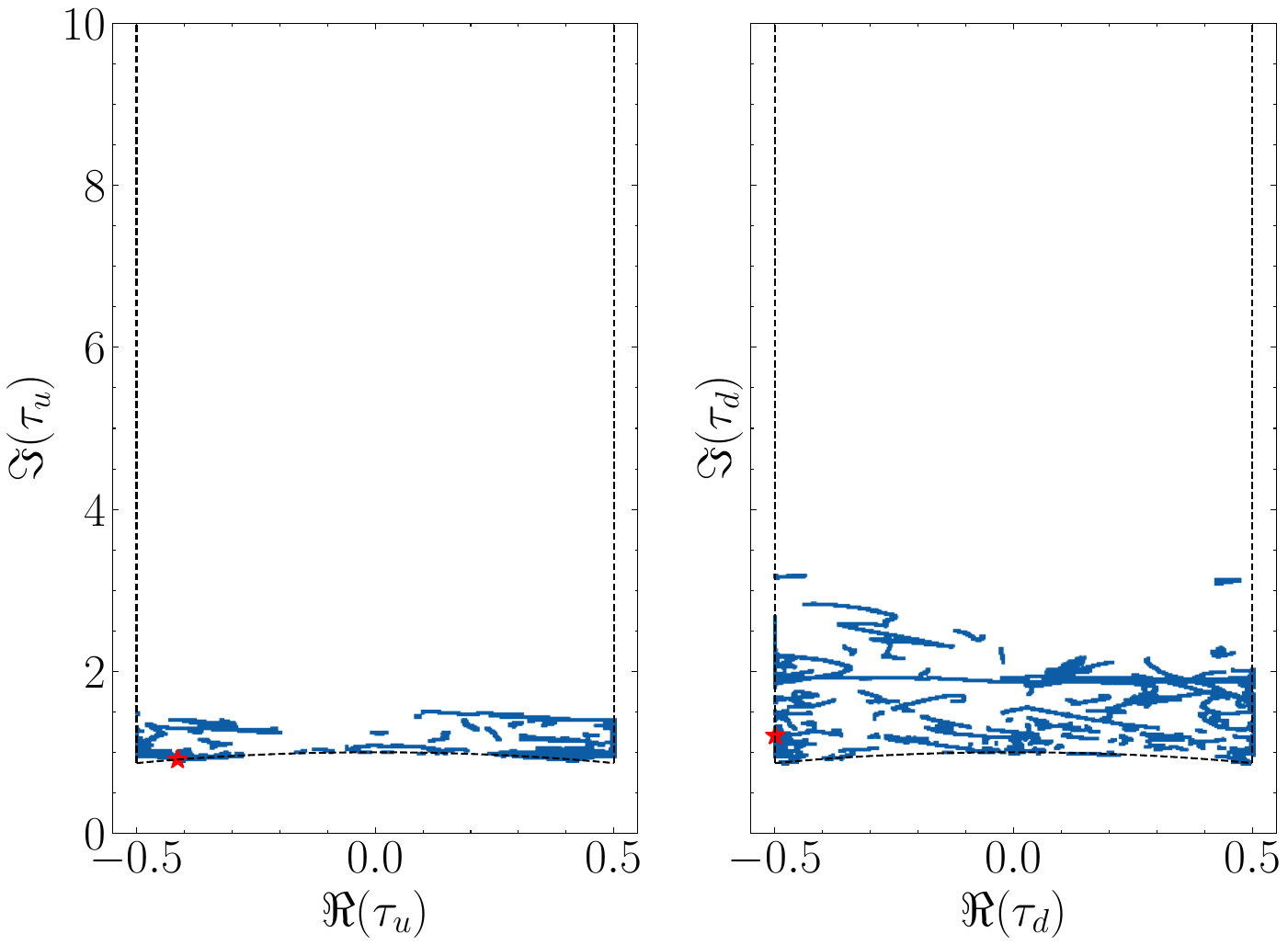}
	\caption{$\tau_u$ and $\tau_d$ values for the CMAES scan. All the points hold predictions within 3-$\sigma$. The red star point represents the best fit point,~\cref{eq:best_point}. Dashed line represents the boundary of the fundamental domain. }
	\label{fig:cmaes_taus}
\end{figure}

\begin{figure}[H]
	\centering
	\includegraphics[width=0.666\textwidth]{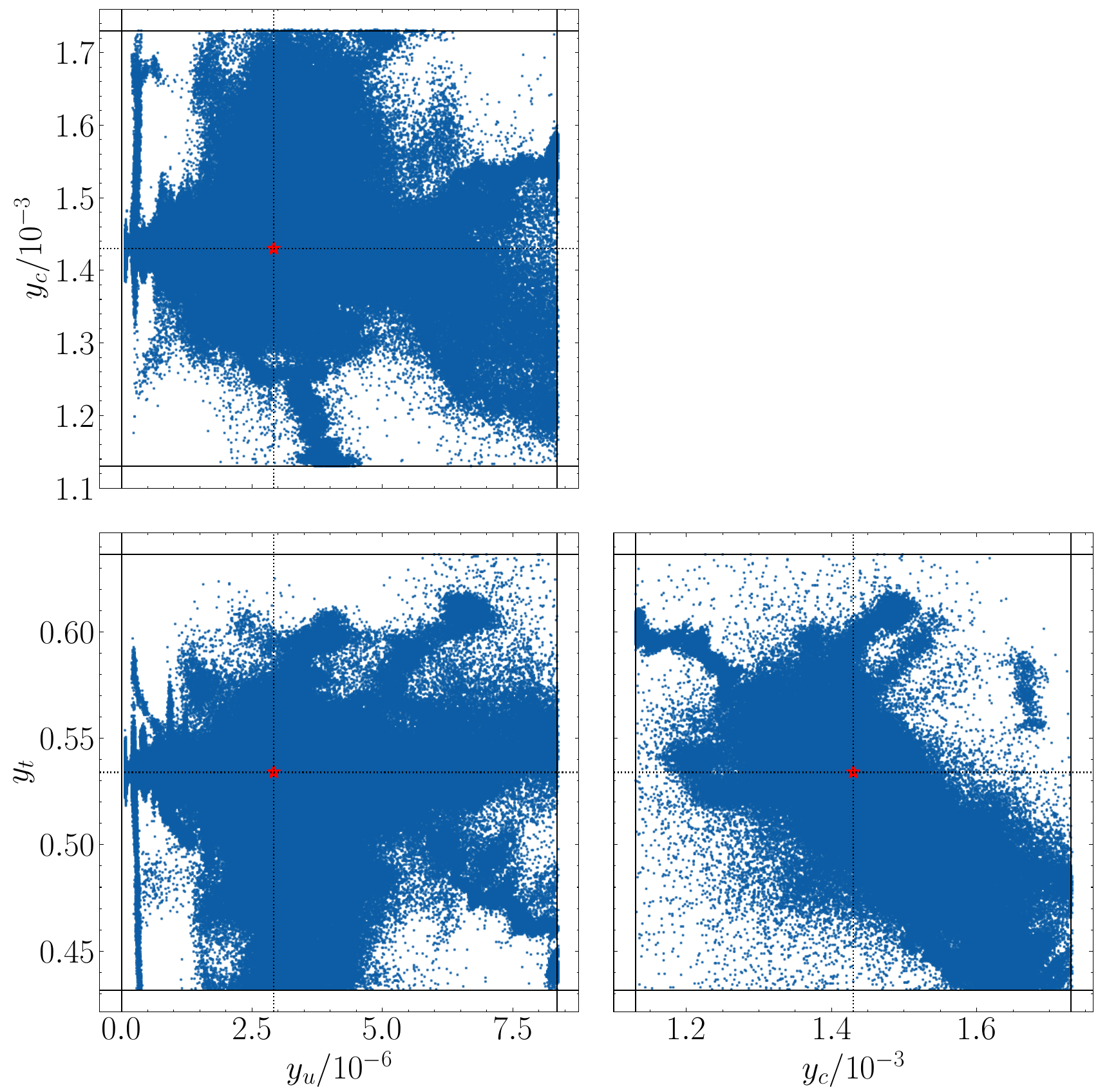}
	\caption{Up-type quark Yukawa eigenvalues obtained for the CMAES scan. All the points hold predictions within 3-$\sigma$. The red star point represents the best fit point,~\cref{eq:best_point}. The dashed (full) lines represent the central value (3-$\sigma$ bounds) from~\cref{tab:quarkdata}.}
	\label{fig:cmaes_up_yukawa}
\end{figure}

\begin{figure}[H]
	\centering
	\includegraphics[width=0.666\textwidth]{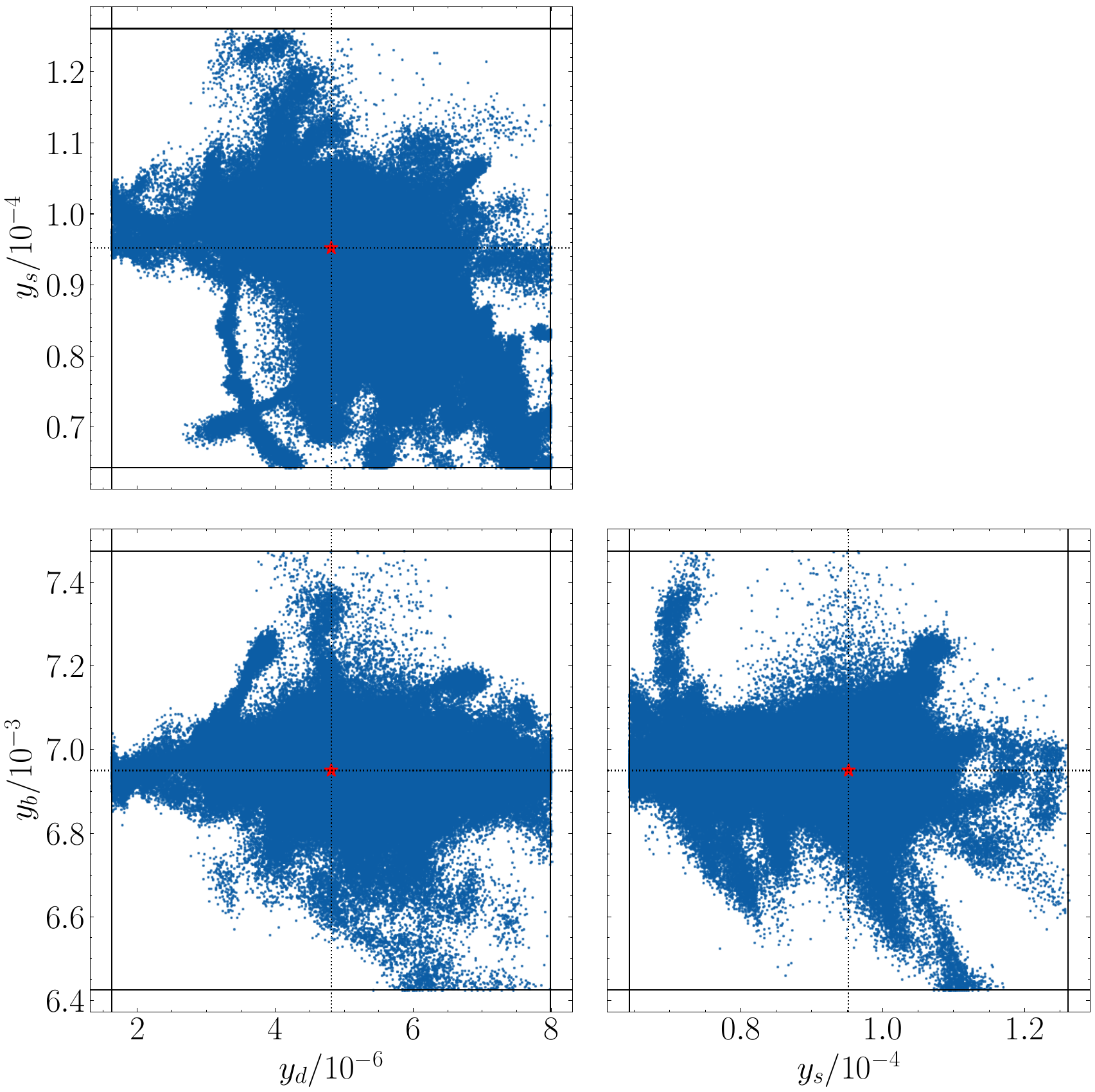}
	\caption{Down-type quark Yukawa eigenvalues obtained for the CMAES scan. All the points hold predictions within 3-$\sigma$. The red star point represents the best fit point,~\cref{eq:best_point}. The dashed (full) lines represent the central value (3-$\sigma$ bounds) from~\cref{tab:quarkdata}.}
	\label{fig:cmaes_down_yukawa}
\end{figure}

\begin{figure}[H]
	\centering
	\includegraphics[width=1\textwidth]{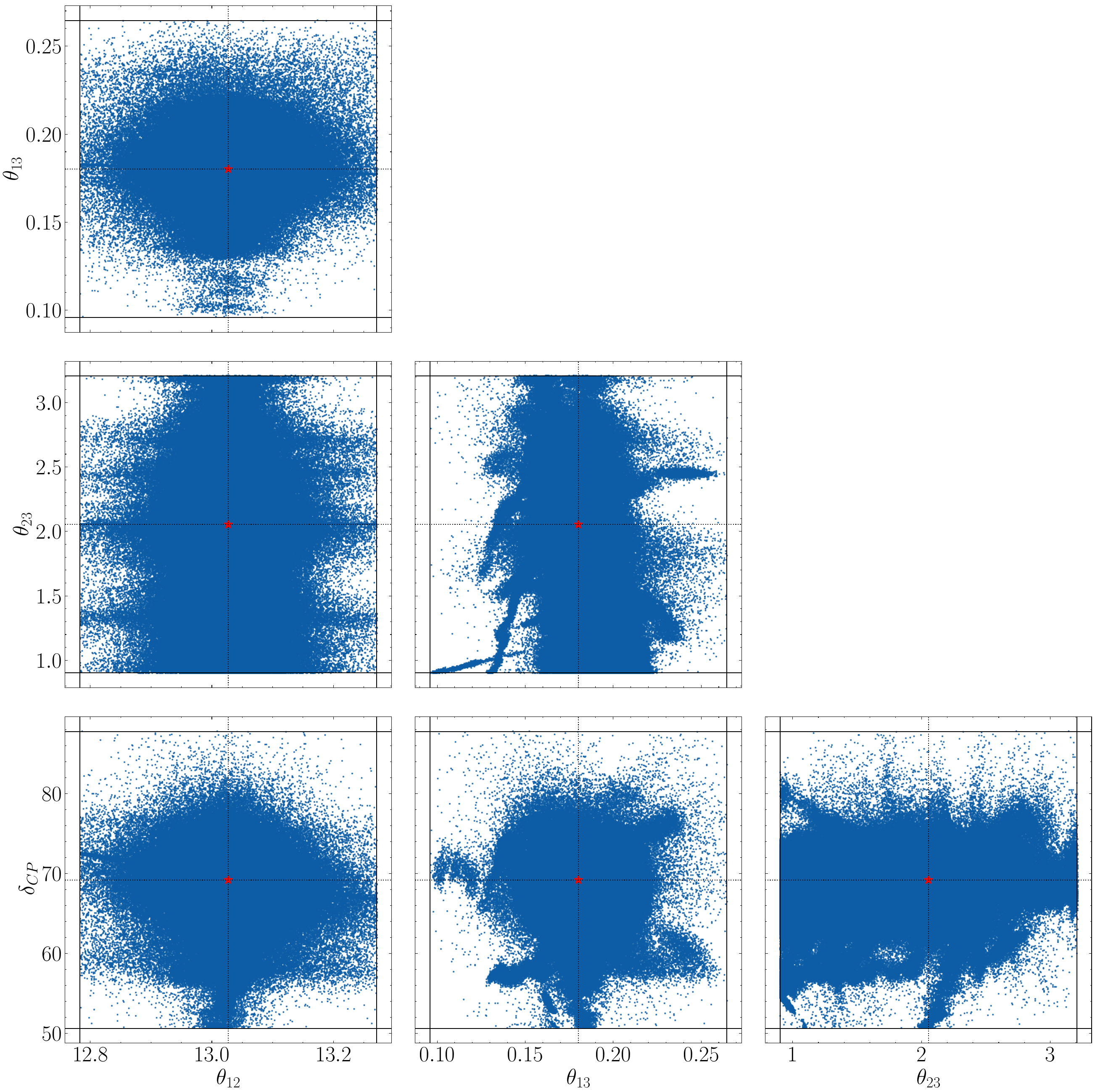}
	\caption{CKM angles and CP phase obtained for the CMAES scan. All the points hold predictions within 3-$\sigma$. The red star point represents the best fit point,~\cref{eq:best_point}. The dashed (full) lines represent the central value (3-$\sigma$ bounds) from~\cref{tab:quarkdata}.}
	\label{fig:cmaes_angles}
\end{figure}

\begin{figure}[H]
	\centering
	\includegraphics[width=0.666\textwidth]{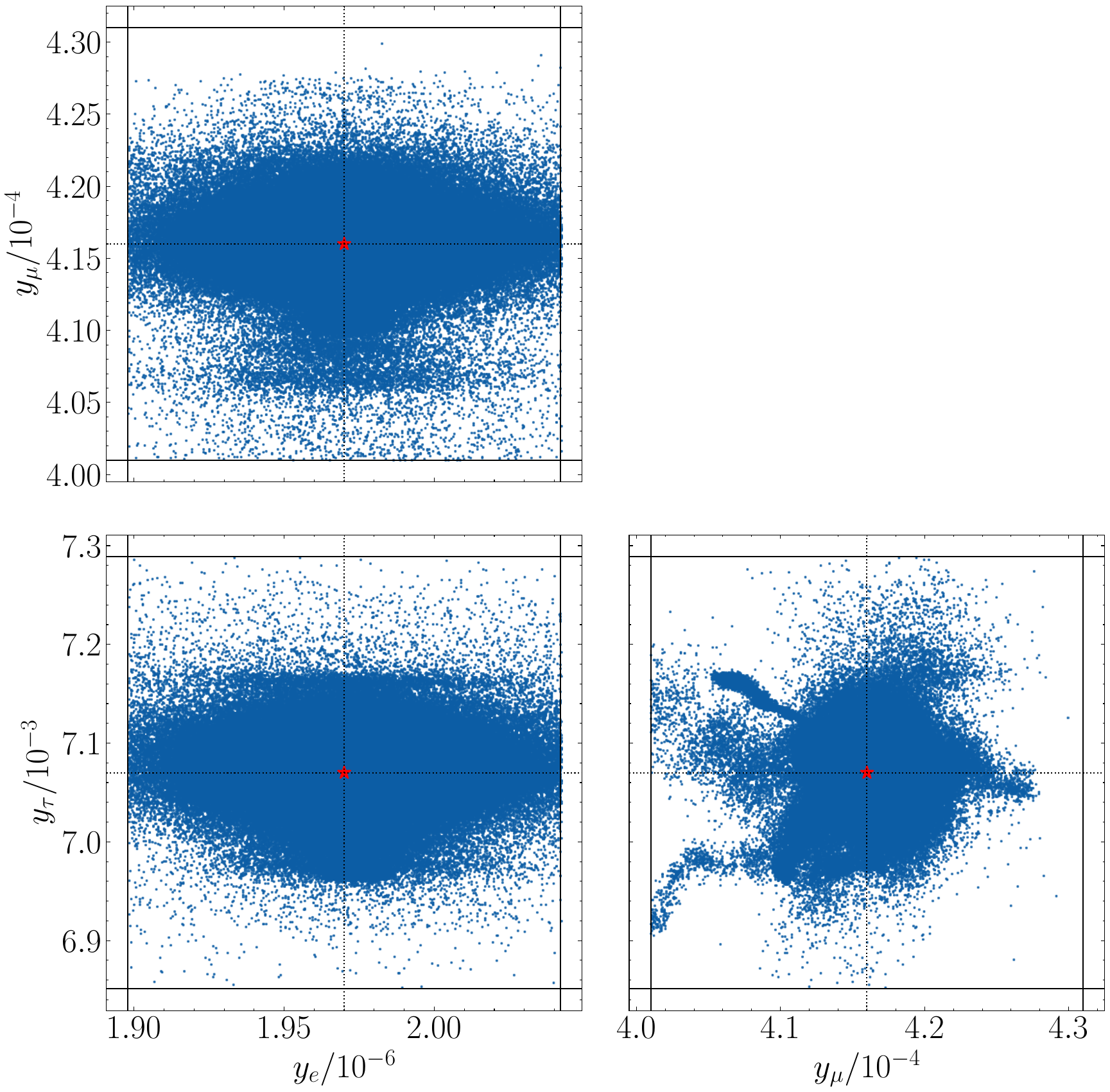}
	\caption{Charged leptons Yukawa eigenvalues obtained for the CMAES scan. All the points hold predictions within 3-$\sigma$. The red star point represents the best fit point,~\cref{eq:best_point}. The dashed (full) lines represent the central value (3-$\sigma$ bounds) from~\cref{tab:leptondata}.}
	\label{fig:cmaes_charged_leptons_yukawa}
\end{figure}

\begin{figure}[H]
	\centering
	\includegraphics[width=0.333\textwidth]{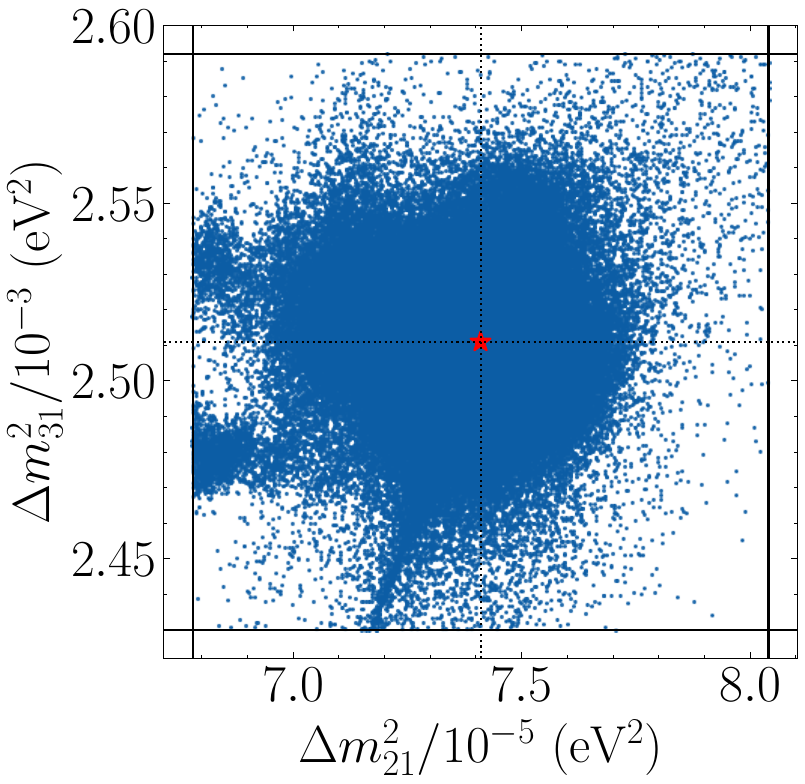}
	\caption{Neutrino squared mass differences obtained for the CMAES scan. All the points hold predictions within 3-$\sigma$. The red star point represents the best fit point,~\cref{eq:best_point}. The dashed (full) lines represent the central value (3-$\sigma$ bounds) from~\cref{tab:leptondata}.}
	\label{fig:cmaes_neutrino_squared_mass_diff}
\end{figure}

\begin{figure}[H]
	\centering
	\includegraphics[width=1\textwidth]{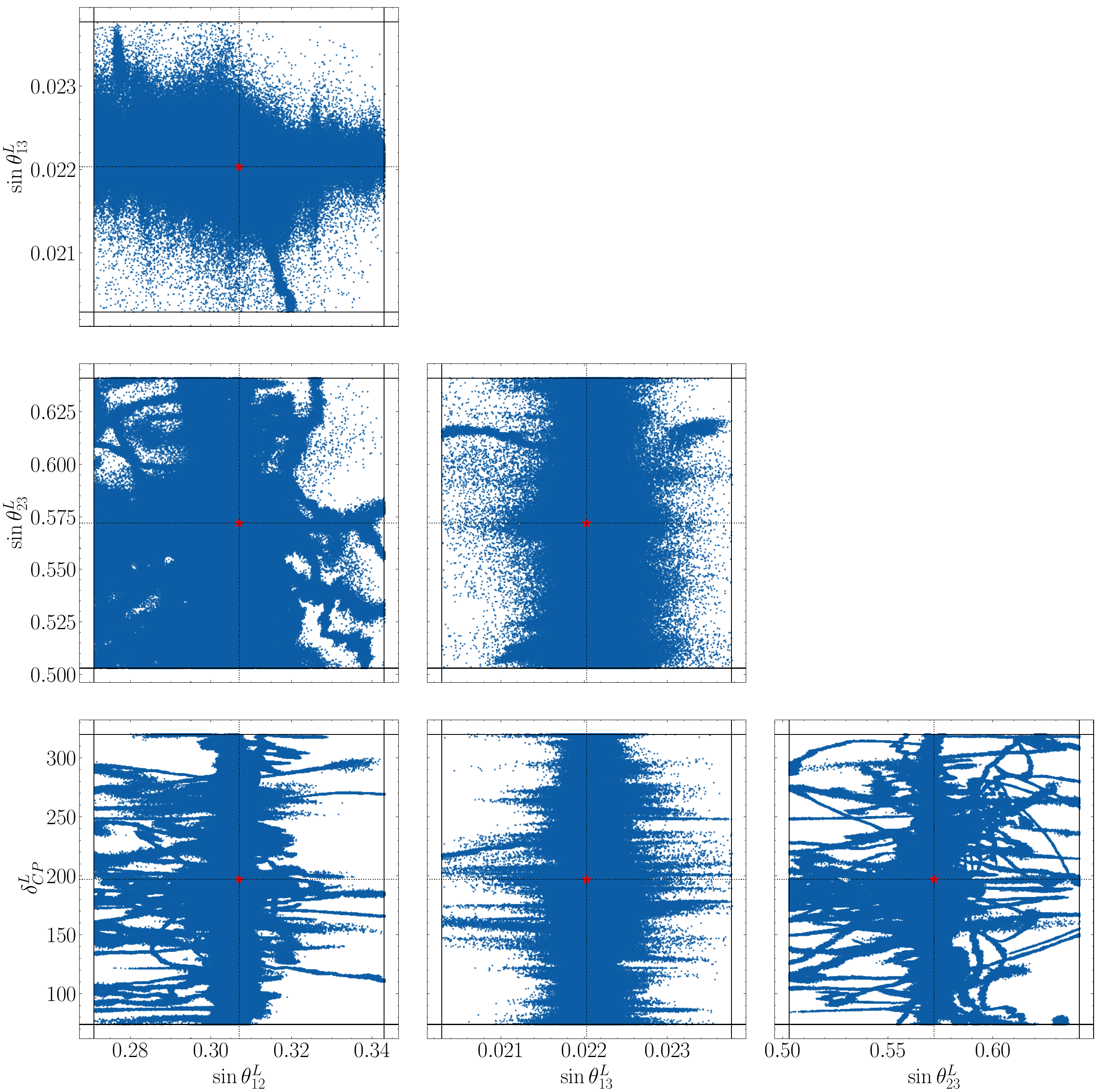}
	\caption{PMNS angles and CP phase obtained for the CMAES scan. All the points hold predictions within 3-$\sigma$. The red star point represents the best fit point,~\cref{eq:best_point}. The dashed (full) lines represent the central value (3-$\sigma$ bounds) from~\cref{tab:leptondata}.}
	\label{fig:cmaes_PMNS_angles}
\end{figure}

In~\cref{fig:cmaes_taus} we show the values of the moduli field that were obtained by CMAES, where we see that lower values of the imaginary part of the moduli are preferred, and most points have $\Im(\tau_i)\lesssim 2$. We omit scatter plots for the remaining parameters as these are, in principle, computable in F-Theory, and the details of their numerical realisation are left to future study. We also note that one should not attempt to make statistical interpretations of the results of CMAES, as it is not an algorithm designed to populate a posterior (as Monte Carlo Markov Chains do in Bayesian inference) as it produces points through the path of quickest descent of the loss function (and therefore the points should also not be used for frequentist interpretations as one usually does with random sampling). However, all points are within 3-$\sigma$ of all observables and therefore have a very high likelihood, or, conversely, a very small $\chi^2$.
We first look at the results pertaining to the quark data. In~\cref{fig:cmaes_up_yukawa} we can observe the resulting values for the up-type quark Yukawa eigenvalues of points obtained, and in~\cref{fig:cmaes_down_yukawa} we present the equivalent plots for the down-type quarks. We see that many points can be arbitrarly close to the central value, but also span the region within the 3-$\sigma$ limits, showing that the model produces a good fit to the data. The same can be observed in~\cref{fig:cmaes_angles} for the CKM mixing angle and CP violating phase.
In~\cref{fig:cmaes_charged_leptons_yukawa,fig:cmaes_neutrino_squared_mass_diff,fig:cmaes_PMNS_angles} we can arrive at similar conclusions regarding the charged lepton Yukawa eigenvalues, neutrinos squared mass differences, PMNS mixing angles, and CP violating phase.

The above results show that our model can fit the data very well, with the best point having a likelihood close to unity or, conversely, a vanishingly small $\chi^2$. However, the problem is over-parametrised by the number of superpotential coefficients, which, although in principle calculable in F-Theory, are considered free parameters in this analysis. To assess whether we can reduce the parametric freedom, we considered alternative scenarios with reduced parametric freedom with respect to the moduli. In our first alternative scenario, we fixed the moduli to take the same values (i.e. $\tau_u=\tau_d$ but otherwise allowed the moduli to take values in the fundamental domain~\cref{eq:domain}) even though our Fluxed GUT naturally provides distinct moduli for each Yukawa type. The scans converged successfully as before, from which we can conclude that our model does not require two independent moduli to fit the data.

For the second case, we fixed the moduli to special values $\tau_u,\ \tau_d \in \{ i,\ i\infty, \omega = \exp{(2\pi i/3)}\}$ (but not necessarily equal). In this scenario, CMAES failed to find points that fit the data. To further study this scenario, we restricted the problem to only fit the quark data, and even then the best-case scenario was for the configuration $\tau_u =i \infty$, $\tau_d=i$, for which we were able to fit all the observables within 3-$\sigma$ except for the $\theta_{12}$ angle of the CKM matrix. The fact that the best-case scenario relies on $\tau_u =i \infty$ suggests that it is indeed not possible to find good points that fit all the data with the moduli stabilised at special values, as we have seen in~\cref{fig:cmaes_taus} that the scans showed a preference for small values of $\Im(\tau_i)$.

Therefore, we conclude that, despite being over-parameterised, the model works with fewer parameters although we lack F-theoretical motivations to restrict their number. The realisation of a model with fewer parameters within the framework presented in this paper is left to future work, as well as the computation of the superpotential coefficients.

\section{Summary and Conclusions\label{sec:conclusions}}

In summary, after reviewing modular symmetry and Yukawa couplings in F-theory,
we have discussed the possible origin of discrete modular symmetries in local F-Theory constructions and applied the conclusions to an explicit F-Theory inspired bottom-up model.
The key point is that in local F-Theory the Yukawa couplings should inherit modular symmetry properties due the dependence of the holomorphic functions on the complex structure moduli.

Having argued that  finite modular family symmetry should emerge from local F-theory inspired Fluxed GUTs, we then pursued a bottom-up approach to such constructions that share important similarities and features to known F-Theory models.
As a concrete example, we have analysed in detail a concrete bottom-up F-Theory inspired fluxed $SU(5)$ GUT with modular $S_4$ family symmetry. The model fits the fermion mass and mixing data very well and serves as a demonstration of the bottom-up approach to modular family symmetry in Fluxed GUTs. It is the first attempt for implementing a discrete modular family symmetry in Fluxed GUTs model building.

The example discussed is an extension of a model already presented in~\cite{Callaghan:2011jj}, but with important differences. The fields here transform as a representation of a discrete modular family group, in our case $S_4$, and also carry modular weights. The Yukawa couplings, arising from the overlapping wavefunction integrals at matter curve intersections, are promoted to modular forms of appropriate weight. We show that the model fits the data very well, with the best point having vanishing $\chi^2$, although it is over-parametrised with respect to superpotential coefficients, which while in principle computable in F-Theory are left as free parameters in the analysis presented in this work.

Although this approach is promising, we have identified aspects that need further detailed exploration, such as the explicit construction of global F-Theory models where the matter curve's dependence on the complex structure moduli will become apparent. Such a reformulation would unravel the modular symmetric properties of the holomorphic function leading to a unique computation of the Yukawa couplings. We leave these challenging issues for future work.

In conclusion, we have shown that finite modular family symmetry should emerge from local bottom-up Fluxed GUTs constructions inspired by F-Theory, opening up new avenues of model building in this rich framework. Additionally, our construction suggests new areas of research into the origin of family symmetries in F-Theory.

\section*{Acknowledgements}
MCR is supported by the STFC under Grant No.~ST/T001011/1. SFK acknowledges the STFC Consolidated Grant ST/L000296/1 and the European Union's Horizon 2020 Research and Innovation programme under Marie Sk\l{}odowska-Curie grant agreement HIDDeN European ITN project (H2020-MSCA-ITN-2019//860881-HIDDeN). GKL would like to thank the staff of the Mainz Institute for Theoretical Physics (MITP) for their kind hospitality where part of this work has been carried out.

\newpage

\appendix

\section{Appendix}

\subsection{Definitions of modular forms}
For the modular group $\Gamma_4 \cong S_4$, there are five linearly independent modular forms of the lowest non-trivial weight $k=2$, denoted as $Y_i(\tau)$ for $i = 1, 2,..,5$, which form a doublet 2 and a triplet $3^{\prime}$ under the modular $S_4$ symmetry transformations, namely,

\begin{align}
	Y_2^{(2)}=\begin{pmatrix}
		          Y_1(\tau) \\
		          Y_2(\tau)
	          \end{pmatrix},\quad
	Y_{3^{\prime}}^{(2)}=\begin{pmatrix}
		                     Y_3(\tau) \\
		                     Y_4(\tau) \\
		                     Y_5(\tau)
	                     \end{pmatrix}~,
\end{align}

\begin{align}
	Y_1^{(4)}=Y_1^2+Y^2_2,\quad Y_2^{(4)}=\begin{pmatrix}
		                                      Y_2^2-Y_1^2 \\
		                                      2Y_1Y_2
	                                      \end{pmatrix},\; Y_3^{(4)}=\begin{pmatrix}
		                                                                 -2Y_2Y_3              \\
		                                                                 \sqrt{3}Y_1Y_5+Y_2Y_4 \\
		                                                                 \sqrt{3}Y_1Y_4+Y_2Y_5
	                                                                 \end{pmatrix},\; Y_{3^{\prime}}^{(4)}= \begin{pmatrix}
		                                                                                                        2Y_1Y_3               \\
		                                                                                                        \sqrt{3}Y_2Y_5-Y_1Y_4 \\
		                                                                                                        \sqrt{3}Y_2Y_4-Y_1Y_5
	                                                                                                        \end{pmatrix}~,
\end{align}

\begin{align}
	Y_1^{(6)}=Y_1(3Y_2^2-Y_1^2),\quad Y_{1^{\prime}}^{(6)}=Y_2(3Y_1^2-Y_2^2)~,
\end{align}

\begin{align}
	Y_2^{(6)}=(Y_1^2+Y_2^2) \begin{pmatrix}
		                        Y_1 \\
		                        Y_2
	                        \end{pmatrix},\;\;
	Y_3^{(6)}=\begin{pmatrix}
		          Y_1(Y_4^2-Y_5^2)           \\
		          Y_3(Y_1Y_5+\sqrt{3}Y_2Y_4) \\
		          -Y_3(Y_1Y_4+\sqrt{3}Y_2Y_5)
	          \end{pmatrix}~,
\end{align}

\begin{align}
	Y_{3^{\prime},1}^{(6)}=(Y_1^2+Y_2^2)\begin{pmatrix}
		                                    Y_3 \\
		                                    Y_4 \\
		                                    Y_5
	                                    \end{pmatrix},\quad Y_{3^{\prime},2}^{(6)}=\begin{pmatrix}
		                                                                               Y_2(Y_5^2-Y_4^2)            \\
		                                                                               -Y_3(Y_2Y_5-\sqrt{3}Y_1Y_4) \\
		                                                                               Y_3(Y_2Y_4-\sqrt{3}Y_1Y_5)
	                                                                               \end{pmatrix}~,
\end{align}
and
\begin{align}
	Y_{1}^{(8)}=(Y_1^2+Y_2^2)^2,\quad Y_{2,1}^{(8)}=(Y_1^2+Y_2^2)\begin{pmatrix}
		                                                             Y_2^2-Y_1^2 \\
		                                                             2Y_1Y_2
	                                                             \end{pmatrix},\quad Y_{2,1}^{(8)}=(Y_1^2-3Y_2^2)\begin{pmatrix}
		                                                                                                             Y_1^2 \\
		                                                                                                             Y_1Y_2
	                                                                                                             \end{pmatrix}~.
\end{align}

The expressions of modular forms can be derived with the help of the Dedekind $\eta$ function

\begin{align}
	\eta(\eta)=q^{1/24}\Pi_{n=1}^{\infty}(1-q^n), \;\;q=e^{2\pi i \tau},
\end{align}
and its derivative
\begin{align}
	Y(\alpha_1,..,\alpha_6|\tau)\equiv \frac{d}{d\tau}\big[ & \alpha_1\log\eta(\tau+\frac{1}{2})+\alpha_2\log\eta(4\tau)+ \alpha_3\log\eta(\frac{\tau}{4})+\alpha_4\log\eta(\frac{\tau+1}{4})+\notag \\
	                                                        & +\alpha_5\log\eta(\frac{\tau+2}{4})+\alpha_6\log\eta(\frac{\tau+3}{4})\big],
\end{align}
with the coefficients $\alpha_i$ (for $i = 1, 2,··· , 6) $ fulfilling $\alpha_1 +···+ \alpha_6 = 0$. More explicitly, we
have

\begin{align}
	 & Y_1(\tau)\equiv iY(1,1,-1/2,-1/2,-1/2,-1/2|\tau),\notag                           \\
	 & Y_2(\tau)\equiv iY(0,0,\sqrt{3}/2,-\sqrt{3}/2,\sqrt{3}/2,-\sqrt{3}/2|\tau),\notag \\
	 & Y_3(\tau)\equiv iY(1,-1,0,0,0,0|\tau),\notag                                      \\
	 & Y_4(\tau)\equiv iY(0,0,-1/\sqrt{2},i/\sqrt{2},1/\sqrt{2},-i/\sqrt{2}|\tau),\notag \\
	 & Y_5(\tau)\equiv iY(0,0,-1/\sqrt{2},-i/\sqrt{2},1/\sqrt{2},i/\sqrt{2}|\tau)~.
\end{align}

In the above computations, we have used the following Fourier expansions of the modular forms:

\begin{align}
	Y_1 & =-3 \pi  \left(39 q^9+3 q^8+24 q^7+12 q^6+18 q^5+3 q^4+12 q^3+3 q^2+3 q+\frac{1}{8}\right), \notag      \\
	Y_2 & = 3 \sqrt{3} \pi  \sqrt{q} \left(18 q^8+24 q^7+14 q^6+12 q^5+13 q^4+8 q^3+6 q^2+4 q+1\right), \notag    \\
	Y_3 & =\pi  \left(26 q^9+6 q^8-16 q^7+24 q^6-12 q^5+6 q^4-8 q^3+6 q^2-2 q+\frac{1}{4}\right),\notag           \\
	Y_4 & =-\sqrt{2} \pi  \sqrt[4]{q} \left(48 q^8+30 q^7+31 q^6+32 q^5+18 q^4+14 q^3+13 q^2+6 q+1\right), \notag \\
	Y_5 & =-4 \sqrt{2} \pi  q^{3/4} \left(12 q^8+8 q^7+10 q^6+6 q^5+5 q^4+6 q^3+3 q^2+2 q+1\right)~.
\end{align}

\end{document}